\documentclass[iop]{emulateapj}

\usepackage{subfigure}
\usepackage[T1]{fontenc}
\usepackage{graphicx}
\usepackage{hyperref}

\newcommand{\Ha}{H$\alpha$}
\newcommand{\Lha}{$L_{\mathrm{H}\alpha}$}
\newcommand{\ie}{i.e.\ }
\newcommand{\eg}{e.g.\ }

\shorttitle{H$\alpha$ Imaging of Nearby Seyfert Host Galaxies}
\shortauthors{Theios et al.}

\begin{document}

\title{H$\alpha$ Imaging of Nearby Seyfert Host Galaxies}
\author{Rachel L. Theios\altaffilmark{1}\email{rtheios@astro.caltech.edu}}
\author{Matthew A. Malkan and Nathaniel R. Ross\altaffilmark{2}}
\altaffiltext{1}{Current address: Cahill Center for Astronomy and Astrophysics, California Institute of Technology, MS 249-17, Pasadena, CA 91125, USA}
\altaffiltext{2}{Current address: Raytheon Space and Airborne Systems, 2000 E El Segundo Blvd, El Segundo, CA 90245, USA}
\affil{Department of Physics and Astronomy, University of California, Los Angeles, 430 Portola Plaza, Los Angeles, CA 90095, USA}

\begin{abstract}
  
We used narrowband ($\Delta\lambda = 70$ \AA) interference filters with the CCD imaging camera on the Nickel 1.0 meter telescope at Lick Observatory to observe 31 nearby ($z < 0.03$) Seyfert galaxies in the 12 \micron\ Active Galaxy Sample \citep{spinoglio1989}. We obtained pure emission line images of each galaxy, which reach down to a flux limit of $7.3 \times 10^{-15}~\mathrm{erg}~\mathrm{cm}^{-2}~\mathrm{s}^{-1}~\mathrm{arcsec}^{-2}$, and corrected these images for [N~\textsc{ii}] emission and extinction. We separated the \Ha\ emission line from the ``nucleus'' (central 100--1000 pc) from that of the host galaxy. The extended \Ha\ emission is expected to be powered by newly formed hot stars, and indeed correlates well with other indicators of current SFRs in these galaxies: 7.7 \micron\ PAH, far infrared, and radio luminosity. Relative to what is expected from recent star formation, there is a 0.8 dex excess of radio emission in our Seyfert galaxies. The \Ha\ luminosity we measured in the galaxy centers is dominated by the AGN, and is linearly correlated with hard X-ray luminosity. There is, however, an upward offset of 1 dex in this correlation for Seyfert 1s, because their nuclear \Ha\ emission includes a strong additional contribution from the Broad Line Region. We found a correlation between SFR and AGN luminosity. In spite of selection effects, we concluded that the absence of bright Seyfert nuclei in galaxies with low SFRs is real, albeit only weakly significant. We used our measured spatial distributions of \Ha\ emission to determine what these Seyfert galaxies would look like when observed through fixed apertures (\eg a spectroscopic fiber) at higher redshifts. Although all of these Seyfert galaxies would be detectable emission line galaxies at any redshift, most would appear dominated by ($> 67\%$) their H~\textsc{ii} region emission. Only the most luminous AGN ($\mathrm{log}(L_{\mathrm{H}\alpha}) > 41.5~\mathrm{erg}~\mathrm{s}^{-1}$) would still be identified as such at $z\sim0.3$.

\end{abstract}

\section{Introduction}

Almost all observations of active galactic nuclei (AGN) suffer from the fact that the unresolved nonstellar emission is observed in combination with emission from the surrounding host galaxy. The mixing of nonstellar continuum with the starlight of the galaxy (especially its bright bulge) is a classic problem in the field \citep[e.g.][]{mf1983,mo1983}. A no less challenging problem is distinguishing the emission lines powered by the central engine from those powered by hot young stars in the host galaxy \citep{ho1997,tommasin2008}. In principle it is possible to separate the lines and continuum generated by stars from that of the central engine---they have several spectroscopic differences which are bigger than the range found among the two components separately. However, spectroscopic separation becomes increasingly difficult as the AGN contribution becomes weak relative to that of the host galaxy. This is exactly what happens as the active galaxy system is observed through larger and larger apertures (\ie a slit or fiber spectrograph). Alternately, for a fixed observing aperture and other things being equal, the stellar dilution fraction increases with increasing redshift of the active galaxy, out to $z\sim2$. This host dilution issue is relevant for studies that have selected Seyfert galaxies based on optical emission lines, both locally \citep{kauffmann2003} and at higher redshift \citep{juneau2011,yan2011}.

There are two strategies for dealing with the host galaxy dilution problem. The first, taken by most studies of AGN beyond the local Universe, is to restrict consideration only to ``quasars,'' \ie extreme AGN in which the nonstellar nuclear emission strongly dominates over the host galaxy. This, however, ignores the largest population of less luminous (or obscured) AGN, which may in fact be responsible for the majority of black-hole building in the Universe. The other approach, taken in this paper, is to study these more typical AGN---the Seyfert galaxies---by relying on spatially resolved observations. The problem can in principle be solved with integral-field-unit spectroscopic maps of the full extents of nearby Seyfert galaxies. Here we use a simpler alternative, narrowband interference filter imaging of a representative sample of ``common'' Seyfert galaxies, to separate quantitatively the emission lines powered by black hole accretion in their centers from that by young stars throughout the host galaxies.

\begin{deluxetable*}{lcccccccccc}[htb]
\tabletypesize{\scriptsize}
\tablewidth{0pt}
\tablecaption{4th Quarter 12\micron\ Seyfert Targets \label{tab:obs}}
\tablehead{\colhead{Target Name} & \colhead{R. A.} & \colhead{Decl.} & \colhead{z} & \colhead{Type} & \colhead{Date} & \colhead{On Filter} & \colhead{Off Filter} & \colhead{Exposure} & \colhead{Seeing} & \colhead{Weather} \\ 
\colhead{} & \colhead{(h:m:s)} & \colhead{(d:m:s)} & \colhead{} & \colhead{} & \colhead{} & \colhead{(\AA)} & \colhead{(\AA)} & \colhead{(s)} & \colhead{(arcsec)} & \colhead{} } 
\startdata
Mrk 79 & 07:42:32.8 & +49:48:35 & 0.022189 & Sy 1 & 2012 Dec 14 & 6710 & 6649 & 3600 & 2.3 & Light cirrus \\
 &  &  &  &  & 2013 Feb 10 & 6710 &  & 3600 & 2.9 & Light cirrus \\
NGC 2639 & 08:43:38.1 & +50:12:20 & 0.011128 & Sy 1 & 2012 Oct 19 & 6649 &  & 2700 & 1.8 & Clear \\
 &  &  &  &  & 2013 Feb 9 &  & 6520 & 2800 & 2.4 & Clear \\
NGC 3227 & 10:23:30.6 & +19:51:54 & 0.003859 & Sy 1 & 2012 Dec 14 & 6570 & 6520 & 2700 & 1.8 & Light cirrus \\
NGC 4051 & 12:03:30.6 & +44:31:53 & 0.002336 & Sy 1 & 2013 Feb 9 & 6570 & 6520 & 3100 & 2.4 & Clear \\
NGC 4151 & 12:10:32.6 & +39:24:21 & 0.003319 & Sy 1 & 2013 Feb 10 & 6570 & 6520 & 4000 & 2.9 & Heavy cirrus \\
NGC 5548 & 14:17:59.5 & +25:08:12 & 0.017175 & Sy 1 & 2013 Jul 3 & 6693 & 6606 & 2700 & 2.0 & Clear \\
NGC 7469 & 23:03:15.6 & +08:52:26 & 0.016317 & Sy 1 & 2012 Oct 7 & 6649 & 6520 & 2700 & 2.8 & Clear \\
NGC 7603 & 23:18:56.6 & +00:14:38 & 0.029524 & Sy 1 & 2012 Oct 8 &  & 6649 & 2700 & 1.7 & Clear \\
 &  &  &  &  & 2012 Oct 19 & 6737 &  & 2700 & 1.4 & Clear \\
NGC 4258 & 12:18:57.5 & +47:18:14 & 0.001494 & Sy 1.9 & 2013 Feb 9 & 6570 & 6520 & 2400 & 2.4 & Light cirrus \\
 &  &  &  &  & 2013 Feb 10 & 6570 & 6520 & 2400 & 2.9 & Light cirrus \\
NGC 4579 & 12:37:43.5 & +11:49:05 & 0.005060 & Sy 1.9 & 2013 Feb 9 & 6606 & 6520 & 3200 & 2.4 & Cirrus \\
NGC 5506 & 14:13:14.9 & -03:12:27 & 0.006181 & Sy 1.9 & 2013 Feb 9 & 6606 & 6520 & 3200 & 2.4 & Cirrus \\
NGC 262 & 00:48:47.1 & +31:57:25 & 0.015034 & Sy 2 & 2012 Oct 7 & 6649 & 6520 & 2700 & 2.2 & Clear \\
NGC 660 & 01:43:02.4 & +13:38:42 & 0.002835 & Sy 2 & 2013 Feb 10 & 6570 & 6520 & 1600 & 2.7 & Light cirrus \\
NGC 1068 & 02:42:40.7 & -00:00:48 & 0.003793 & Sy 2 & 2012 Oct 8 & 6570 & 6520 & 2700 & 1.9 & Clear \\
NGC 1056 & 02:42:48.3 & +28:34:27 & 0.005154 & Sy 2 & 2012 Oct 7 & 6606 & 6520 & 2700 & 2.1 & Clear \\
NGC 1144 & 02:55:12.2 & -00:11:01 & 0.028847 & Sy 2 & 2012 Oct 19 & 6737 & 6649 & 2700 & 2.0 & Clear \\
NGC 1194 & 03:03:49.1 & -01:06:13 & 0.013596 & Sy 2 & 2012 Oct 7 & 6649 & 6520 & 2700 & 2.3 & Clear \\
NGC 1241 & 03:11:14.6 & -08:55:20 & 0.013515 & Sy 2 & 2012 Oct 19 & 6649 & 6520 & 2700 & 1.7 & Clear \\
NGC 1320 & 03:24:48.7 & -03:02:32 & 0.008883 & Sy 2 & 2012 Oct 19 & 6606 & 6520 & 2700 & 1.6 & Clear \\
NGC 1667 & 04:48:37.1 & -06:19:12 & 0.015167 & Sy 2 & 2012 Oct 19 & 6649 & 6520 & 2700 & 1.9 & Clear \\
NGC 3079 & 10:01:57.8 & +55:40:47 & 0.003723 & Sy 2 & 2012 Dec 14 & 6570 & 6520 & 2700 & 1.8 & Light cirrus \\
NGC 4501 & 12:31:59.1 & +14:25:13 & 0.007609 & Sy 2 & 2013 Feb 10 & 6606 & 6520 & 3200 & 2.9 & Heavy cirrus \\
NGC 4941 & 13:04:13.1 & -05:33:06 & 0.003696 & Sy 2 & 2013 Feb 10 &  & 6520 & 2400 & 2.9 & Light cirrus \\
 &  &  &  &  & 2014 May 6 & 6570 &  & 4500 & 2.8 & Clear \\
NGC 5929 & 15:26:06.1 & +41:40:14 & 0.008312 & Sy 2 & 2014 May 20 & 6606 & 6520 & 2800 & 2.6 & Light cirrus \\
NGC 5953 & 15:34:32.4 & +15:11:38 & 0.006555 & Sy 2 & 2014 May 6 &  & 6520 & 4500 & 2.8 & Clear \\
 &  &  &  &  & 2014 May 20 & 6606 &  & 2600 & 2.6 & Light cirrus \\
NGC 6574 & 18:11:51.2 & +14:58:54 & 0.007612 & Sy 2 & 2013 Jul 17 & 6606 & 6520 & 2700 & 1.6 & Clear \\
NGC 7674 & 23:27:56.7 & +08:46:45 & 0.028914 & Sy 2 & 2012 Oct 19 & 6737 & 6606 & 2700 & 1.5 & Clear \\
Arp 220 & 15:34:57.1 & +23:30:11 & 0.018126 & LINER & 2013 Jul 3 & 6693 & 6606 & 2700 & 2.0 & Clear \\
NGC 6384 & 17:31:24.3 & +07:03:37 & 0.005554 & LINER & 2013 Jul 16 & 6606 & 6520 & 2700 & 1.8 & Clear \\
NGC 6670 & 18:33:35.4 & +59:53:20 & 0.028853 & non-Sy & 2014 May 21 & 6737 & 6606 & 2100 & 1.9 & Clear \\
NGC 6764 & 19:08:16.4 & +50:56:00 & 0.008059 & LINER & 2013 Jul 16 & 6606 & 6520 & 2700 & 1.4 & Clear \\
\enddata
\tablecomments{ Exposure times refer to total integration times in a single narrowband filter, thus the total integration time for an object is twice this value.}
\end{deluxetable*}

Even with mediocre, uncorrected ground-based seeing, our imaging is adequate to separate the central few hundred parsecs of Seyfert nuclei at $z\sim0.01$ from their host galaxies. This is sufficient for our purposes, since nearly all of the AGN-powered line emission (from the Broad Line Region, BLR, and the Narrow Line Region, NLR) is confined to the central 100 parsecs, where it usually dominates over line emission from H~\textsc{ii} regions in the host galaxy. The results we obtain for these nearby Seyfert galaxies can then be extended to predict how the same set of galaxies would appear if they were observed at higher redshift, with much coarser spatial resolution.

\section{Observations and Data}

We selected 27 Seyfert galaxies and four LINERs/non-Seyferts from the Extended 12 \micron\ Seyfert Sample of \citet{rush1993} visible from the northern hemisphere, and with redshift less than 0.03. We observed these galaxies on twelve nights between October 2012 and May 2014 (see Table \ref{tab:obs}) using the narrowband filters on the Nickel 40-inch telescope at Lick Observatory. We used the Nickel Direct Imaging Camera (CCD-C2), which has $2048 \times 2048$ pixels, read out with $2 \times 2$ binning to yield $1024 \times 1024$ pixels 0.37 arcseconds on a side. We summarize weather and seeing conditions for each galaxy observation in Table \ref{tab:obs}. Typical conditions were clear with little to no moonlight and seeing of about two arcseconds FWHM. For each galaxy, we used one narrowband filter centered as closely as possible on \Ha\ to measure the flux of that line, and another narrowband filter offset from \Ha\ to measure the underlying continuum. We selected the continuum filter for each galaxy to avoid contamination from other emission lines. The filters we used were (central wavelength/FWHM in Angstroms): 6520/75, 6570/70, 6606/75, 6649/76, 6693/76, 6710/100, 6737/76. In general, we obtained three exposures per galaxy in each of the \Ha\ and continuum filters, with individual exposure times ranging from 700 to 1500 seconds, depending on the conditions, in order to build up enough background counts to avoid being read-noise limited. We dithered the telescope between exposures in order to mitigate the effect of hot pixels and several bad columns on the detector. We obtained bias and twilight sky flat field frames each night for each filter used.

\begin{figure*}[htb]
\begin{minipage}{180mm}	
  \centering
  \subfigure{
    \includegraphics[width=0.3\textwidth]{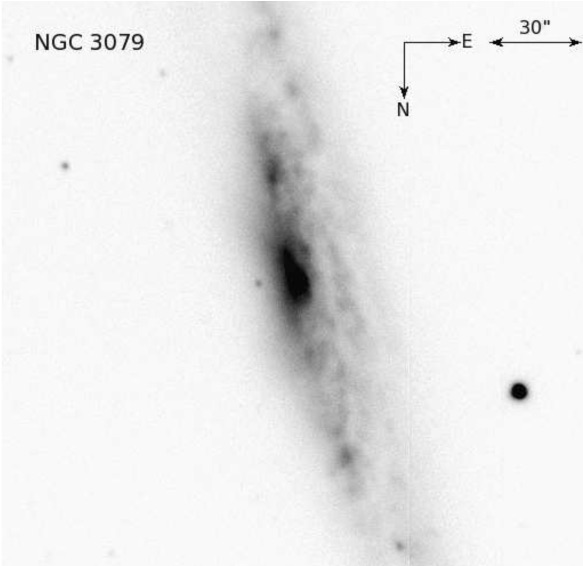}
  }
  \subfigure{
    \includegraphics[width=0.3\textwidth]{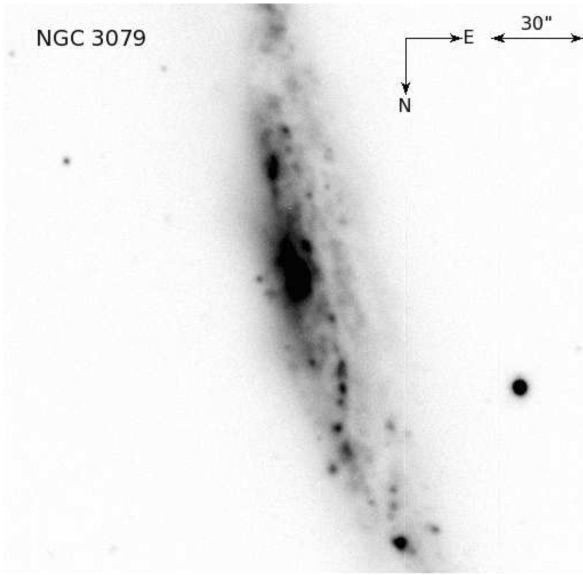}
  }
  \subfigure{
    \includegraphics[width=0.3\textwidth]{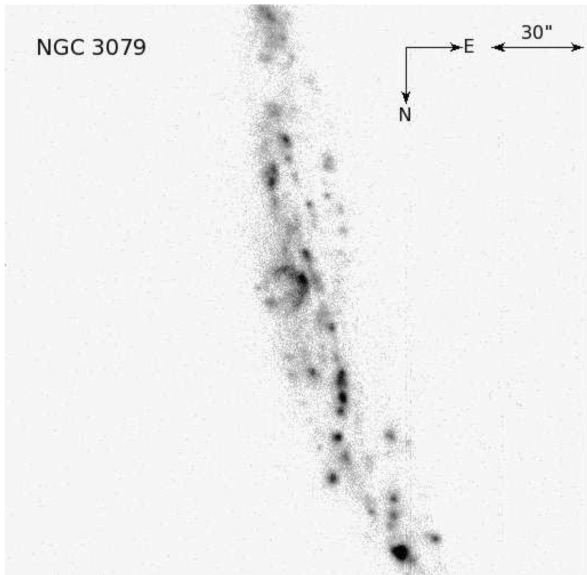}
  }
  \subfigure{
    \includegraphics[width=0.3\textwidth]{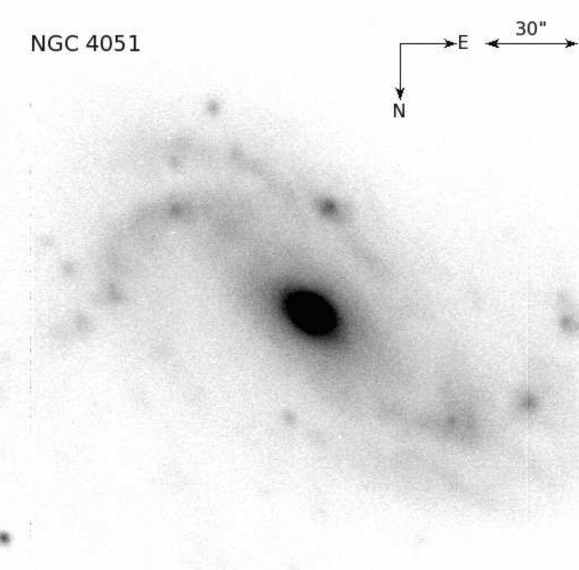}
  }
  \subfigure{
    \includegraphics[width=0.3\textwidth]{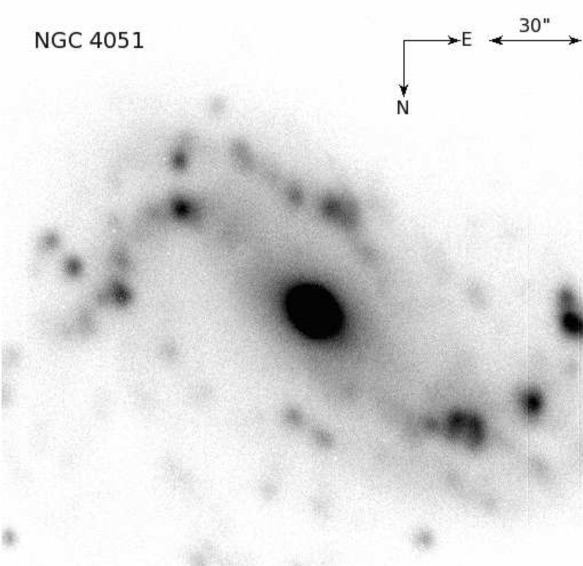}
  }
  \subfigure{
    \includegraphics[width=0.3\textwidth]{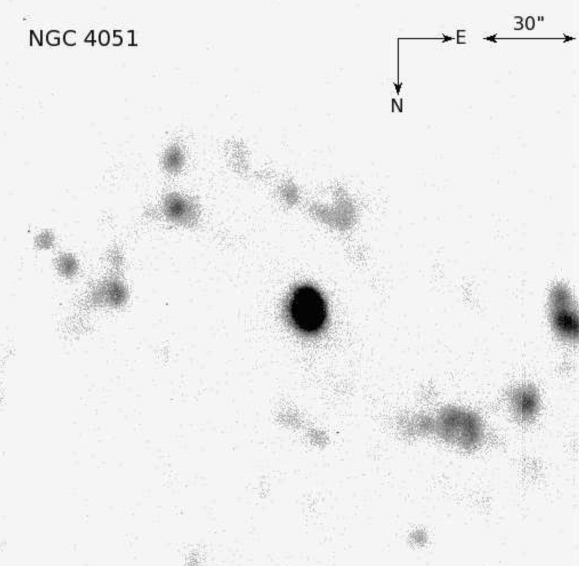}
  }
  \subfigure{
    \includegraphics[width=0.3\textwidth]{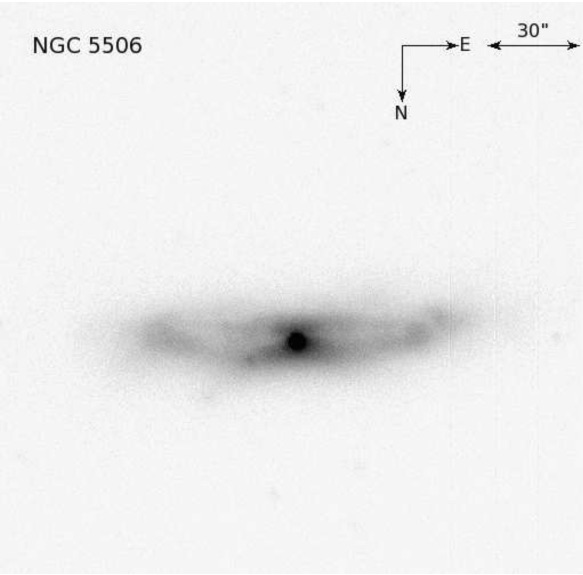}
  }
  \subfigure{
    \includegraphics[width=0.3\textwidth]{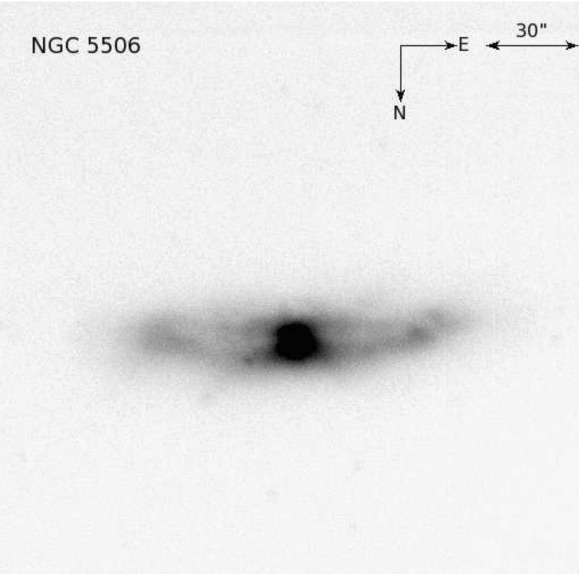}
  }
  \subfigure{
    \includegraphics[width=0.3\textwidth]{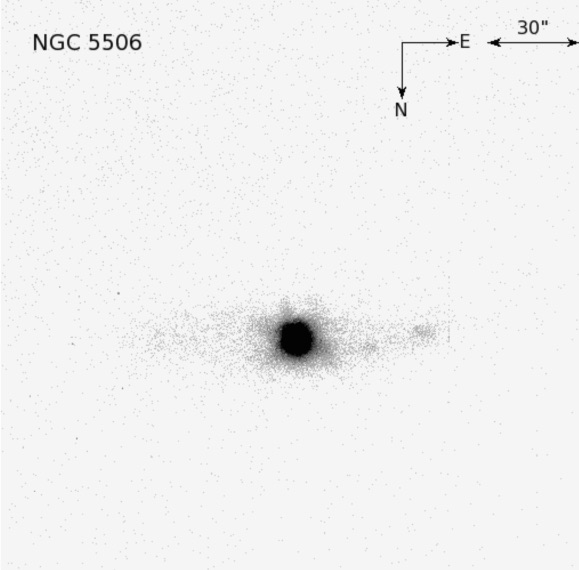}
  }
  \caption{Typical reduction steps for a galaxy. The left column shows the off-band continuum image, the middle is the on-band image, and the right is the continuum-subtracted \Ha\ + [N~\textsc{ii}] image. The scale bar corresponds to 30\arcsec.\label{fig:images}}
  \end{minipage}
  \end{figure*}

We reduced the data using standard IRAF procedures, including bias and flat-field correction. We averaged the three dithered frames in each filter, and subtracted the off-band from the on-band to obtain an image purely in \Ha\ + [N~\textsc{ii}]. Figure \ref{fig:images} gives examples of averaged on-band, off-band continuum, and continuum-subtracted \Ha\ + [N~\textsc{ii}] images. With the IRAF \emph{phot} module, we obtained circular aperture photometry for the continuum-subtracted images and used a flux calibration factor to measure the \Ha\ flux through each aperture.

We photometrically calibrated the narrowband filters using Sloan Digital Sky Survey \citep[SDSS;][]{ahn2012} data. The photometric calibration was performed using the IRAF \emph{phot} task with an aperture 2\farcs 9 in diameter. We used SDSS $r$-band magnitudes to calculate the night constant for selected stars. We then compared the night constant of each star to its SDSS $g-i$ color, and determined that the sensitivity of our filters did not have a significant color dependence. In each filter, the night constants of individual stars showed an RMS 1-$\sigma$ scatter of approximately 5\%. To compare the sensitivity between filters, we measured the flux of the same stars in several different filters. The fluxes in the 6606, 6649, 6520, and 6570~\AA\ filters were consistent with each other to within 5\%, and the 6737~\AA\ filter was consistent with the others to within 20\%. The 6693 \AA\ filter was approximately one magnitude less sensitive than the others. To account for this difference, we scaled the continuum data of the galaxies observed in 6693~\AA\ down by a factor of 0.43.

\begin{figure*}[htb]
\begin{minipage}{180mm}
  \centering
  \includegraphics[width=0.75\linewidth]{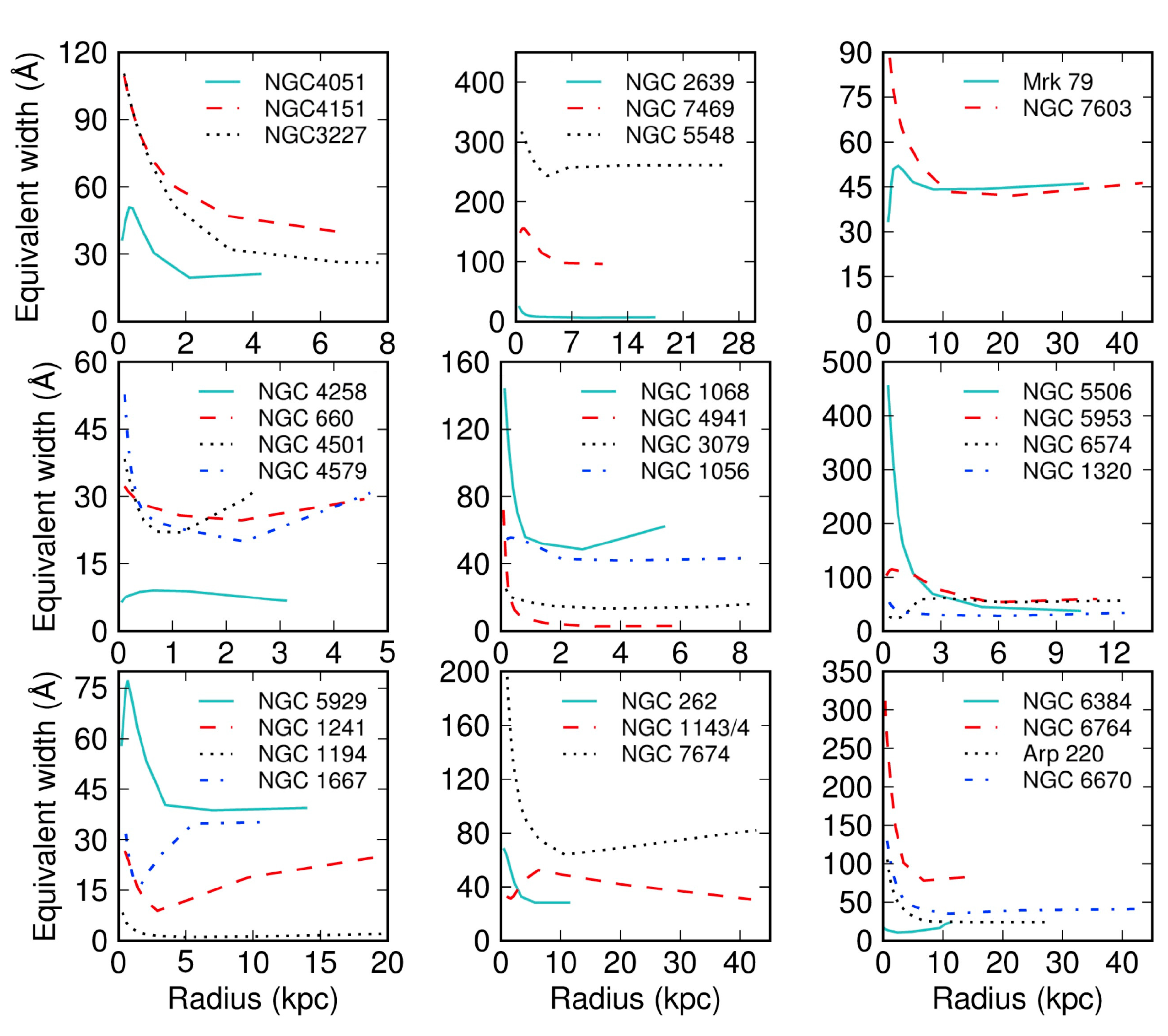}
  \caption{Integrated equivalent width of \Ha\ + [N~\textsc{ii}] versus radius from the galaxy center. The equivalent widths start high in the center, where the emission is dominated by the AGN. The equivalent widths then drop as more of the host galaxy is included; however, in some cases they might rise again due to the presence of star-forming regions. The galaxies in these plots are roughly sorted by distance and Seyfert type. Typical equivalent widths integrated over the whole galaxy range from 20--60 \AA.\label{fig:ew}}
\end{minipage}
\end{figure*}

Since we found that four of the filters were nearly equal in sensitivity, the same flux calibration was used for all galaxies, except for those observed in the 6693 \AA\ filter. Our measurements were made down to a flux limit of $7.3 \times 10^{-15}~\mathrm{erg}~\mathrm{cm}^{-2}~\mathrm{s}^{-1}~\mathrm{arcsec}^{-2}$, which corresponds to 20 counts per pixel for a 900 second exposure. This threshold was selected to exclude background sky fluctuations, while including faint \Ha\ flux from the outer parts of the galaxies. We measured equivalent widths within each aperture over which we performed photometry. Since the \Ha\ + [N~\textsc{ii}] lines were well-centered in the narrowband filter with which they were observed, their equivalent width is defined as: $$ W_{\mathrm{H}\alpha} = \frac{F_{\mathrm{H}\alpha}}{F_\mathrm{c}}\Delta\lambda = 10^{-0.4 \Delta m}\Delta\lambda $$ where $F_{\mathrm{H}\alpha}$ is the flux from the continuum-subtracted image, $F_\mathrm{c}$ is the flux from the continuum, $\Delta m$ is the difference between the on- and off-band magnitudes, and $\Delta\lambda$ is the bandwidth of the filter. Both fluxes were integrated from $r = 0$ out to the aperture radius of interest, which we then converted to a physical radius (kpc) using Virgo-infall-corrected distances from NED\footnote{The NASA/IPAC Extragalactic Database (NED) is operated by the Jet Propulsion Laboratory, California Institute of Technology, under contract with the National Aeronautics and Space Administration.}. Figure \ref{fig:ew} gives the enclosed equivalent width of \Ha\ for each galaxy as a function of radius.

\subsection{Separating the Nuclear Contribution}

\begin{figure}[htb]
  \centering
  \includegraphics[width=\linewidth]{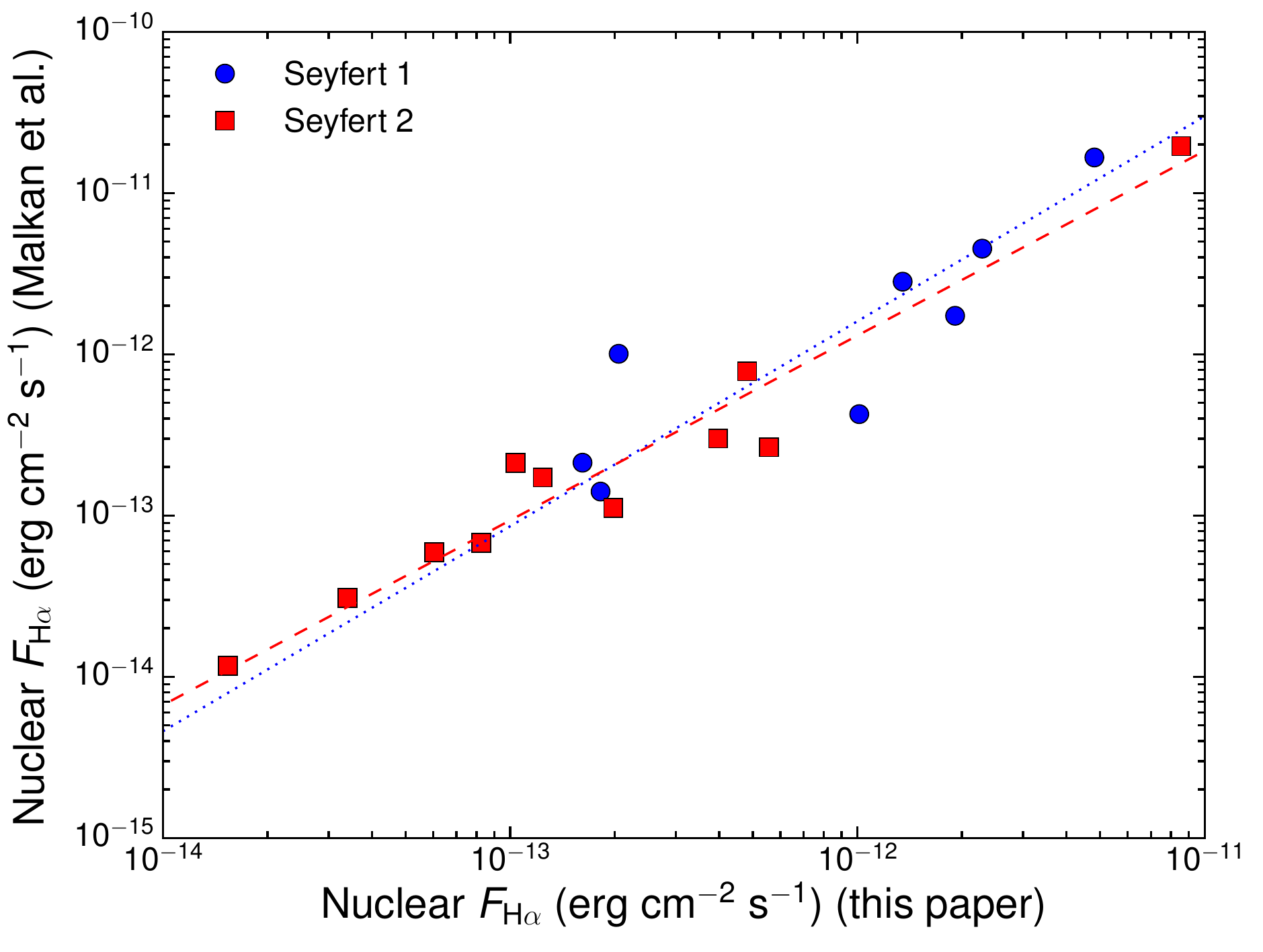}
  \caption{Nuclear \Ha\ flux measured in this paper versus spectroscopic \Ha\ flux from Malkan et al.\ (2016, in preparation). Dotted and dashed lines represent least squares fits of Seyfert 1s and Seyfert 2s respectively. \Ha\ fluxes from this paper are on average 13\% higher than the spectroscopic fluxes.\label{fig:spec}}
\end{figure}

\begin{deluxetable*}{lcccccccc}
\tabletypesize{\small}
\tablewidth{0pt}
\tablecaption{Flux in Selected Apertures \label{tab:flux}}
\tablehead{\colhead{Galaxy} & \colhead{$F_{\mathrm{H}\alpha}$} & \colhead{$F_{\mathrm{H}\alpha}$} & \colhead{$F_{\mathrm{H}\alpha}$} & \colhead{$F_{\mathrm{H}\alpha}$} & \colhead{$F_{\mathrm{H}\alpha}$} & \colhead{$F_{\mathrm{H}\alpha}$} & \colhead{$F_{\mathrm{H}\alpha}$} & \colhead{Type} \\ 
\colhead{} & \colhead{(2\farcs 9)} & \colhead{(3\farcs 7)} & \colhead{(7\farcs 4)} & \colhead{(14\farcs 7)} & \colhead{(37\arcsec)} & \colhead{(74\arcsec)} & \colhead{(147\arcsec)} & \colhead{} } 
\startdata
Mrk 79 & 0.74 & 0.96 & 2.53 & 3.54 & 4.50 & 4.62 & 4.81 & Sy 1 \\
NGC 2639 & 1.09 & 1.32 & 1.88 & 2.40 & 4.14 & 4.46 & 5.03 & Sy 1 \\
NGC 3227 & 11.13 & 13.69 & 19.86 & 24.01 & 28.28 & 31.04 & 35.76 & Sy 1 \\
NGC 4051 & 1.23 & 1.81 & 5.08 & 9.45 & 10.90 & 12.38 & 20.19 & Sy 1 \\
NGC 4151 & 18.02 & 24.53 & 50.81 & 66.09 & 70.57 & 70.88 & 72.04 & Sy 1 \\
NGC 5548 & 7.37 & 8.94 & 12.66 & 15.91 & 19.67 & 19.92 & 19.97 & Sy 1 \\
NGC 7603 & 7.30 & 8.12 & 9.45 & 10.17 & 10.64 & 10.94 & 12.05 & Sy 1 \\
NGC 7469 & 7.98 & 10.93 & 21.24 & 26.67 & 28.17 & 28.44 & 28.56 & Sy 1 \\
NGC 4258 & 0.39 & 0.59 & 2.00 & 4.85 & 14.20 & 27.43 & 45.62 & Sy 1.9 \\
NGC 4579 & 3.35 & 4.44 & 8.46 & 13.44 & 25.12 & 31.40 & 40.32 & Sy 1.9 \\
NGC 5506 & 2.88 & 3.97 & 7.92 & 10.79 & 13.34 & 14.27 & 15.33 & Sy 1.9 \\
NGC 262 & 0.61 & 0.86 & 1.91 & 2.49 & 2.54 & 2.58 & 3.48 & Sy 2 \\
NGC 660 & 0.32 & 0.48 & 1.48 & 3.45 & 8.56 & 14.26 & 22.57 & Sy 2 \\
NGC 1056 & 1.01 & 1.46 & 3.81 & 7.13 & 10.07 & 10.83 & 11.24 & Sy 2 \\
NGC 1068 & 47.39 & 62.16 & 102.23 & 127.47 & 187.45 & 232.04 & 253.91 & Sy 2 \\
NGC 1144 & 0.32 & 0.43 & 1.02 & 3.08 & 7.30 & 8.60 & 9.10 & Sy 2 \\
NGC 1194 & 0.07 & 0.08 & 0.13 & 0.14 & 0.15 & 0.17 & 0.33 & Sy 2 \\
NGC 1241 & 0.50 & 0.65 & 1.12 & 1.24 & 2.78 & 6.92 & 9.57 & Sy 2 \\
NGC 1320 & 1.26 & 1.58 & 2.70 & 4.24 & 6.35 & 7.14 & 8.67 & Sy 2 \\
NGC 1667 & 0.58 & 0.72 & 1.09 & 2.70 & 11.62 & 16.16 & 16.68 & Sy 2 \\
NGC 3079 & 0.22 & 0.31 & 0.85 & 2.33 & 6.46 & 11.75 & 19.39 & Sy 2 \\
NGC 4501 & 1.21 & 1.73 & 4.42 & 8.22 & 18.80 & 46.64 & 86.85 & Sy 2 \\
NGC 4941 & 1.12 & 1.38 & 1.64 & 1.64 & 1.65 & 1.80 & 2.64 & Sy 2 \\
NGC 5929 & 0.71 & 1.02 & 2.76 & 4.98 & 7.27 & 11.70 & 12.38 & Sy 2 \\
NGC 5953 & 0.83 & 1.22 & 3.46 & 8.37 & 16.54 & 17.94 & 45.25 & Sy 2 \\
NGC 6574 & 0.52 & 0.68 & 1.53 & 5.76 & 26.95 & 29.39 & 31.45 & Sy 2 \\
NGC 7674 & 3.22 & 3.63 & 4.47 & 5.87 & 8.89 & 10.24 & 12.13 & Sy 2 \\
NGC 6384 & 0.32 & 0.44 & 1.07 & 2.13 & 3.83 & 7.41 & 13.06 & LINER \\
NGC 6670 & 0.53 & 0.67 & 1.05 & 1.27 & 1.71 & 3.05 & 4.84 & non-Sy \\
NGC 6764 & 4.86 & 5.96 & 9.29 & 12.08 & 13.21 & 14.54 & 18.48 & LINER \\
Arp 220 & 0.21 & 0.29 & 0.63 & 1.22 & 1.64 & 1.64 & 1.64 & LINER \\
\enddata
\tablecomments{Fluxes are in units of $10^{-13}~\mathrm{erg}~\mathrm{cm}^{-2}~\mathrm{s}^{-1}$. Apertures in arcseconds refer to diameters. Fluxes in this table are not corrected for extinction, [N~\textsc{ii}], or aperture losses.}
\end{deluxetable*}

To determine the ``nuclear'' contribution to the \Ha\ flux of each galaxy, we assumed that this emission originated in an unresolved region 2\farcs 9 in diameter. However, due to seeing conditions, some of the nuclear flux spread outside of this aperture. To account for this loss, we determined an aperture correction for each object by comparing the total photometric flux of several stars in the image to the flux within a 2\farcs 9 aperture. We then multiplied the measured nuclear flux by this factor, which ranged between 1.2 and 2.5. In Figure \ref{fig:spec} we compare these seeing-corrected nuclear fluxes to spectroscopic \Ha\ fluxes measured by Malkan et al.\ (2016, in preparation). These quantities show good agreement, and both likely have similar uncertainties. A proper linear least squares fit (the average of the regression of the y-axis onto x and its reverse) of the log-log plot gave a slope of 1.27 for Seyfert 1 galaxies (shown as a blue dotted line) and 1.15 for Seyfert 2 galaxies (red dashed line), with RMS scatters about the best-fit line of 0.25 dex and 0.09 dex for Seyfert 1s and 2s respectively. Our measured fluxes are 13\% higher on average, as might be expected; the spectroscopic slit may not have captured all incident light, especially in the closer galaxies.

These ``nuclear'' fluxes should be dominated by the AGN, because integrated spectra of this central region usually show emission line ratios indicative of nonstellar photoionization (Malkan et al. 2016, in preparation). However, in some of the less luminous AGN, these line ratio diagnostics indicate ``composite'' spectra, which are mixes of AGN and H~\textsc{ii} region lines. Some of the nuclear \Ha\ fluxes therefore may include some small contribution from H~\textsc{ii} regions in and around the galactic nucleus.

\begin{figure*}[htb]
\begin{minipage}{180mm}
  \centering
  \subfigure[Seyfert 1]{
    \includegraphics[width=0.85\linewidth]{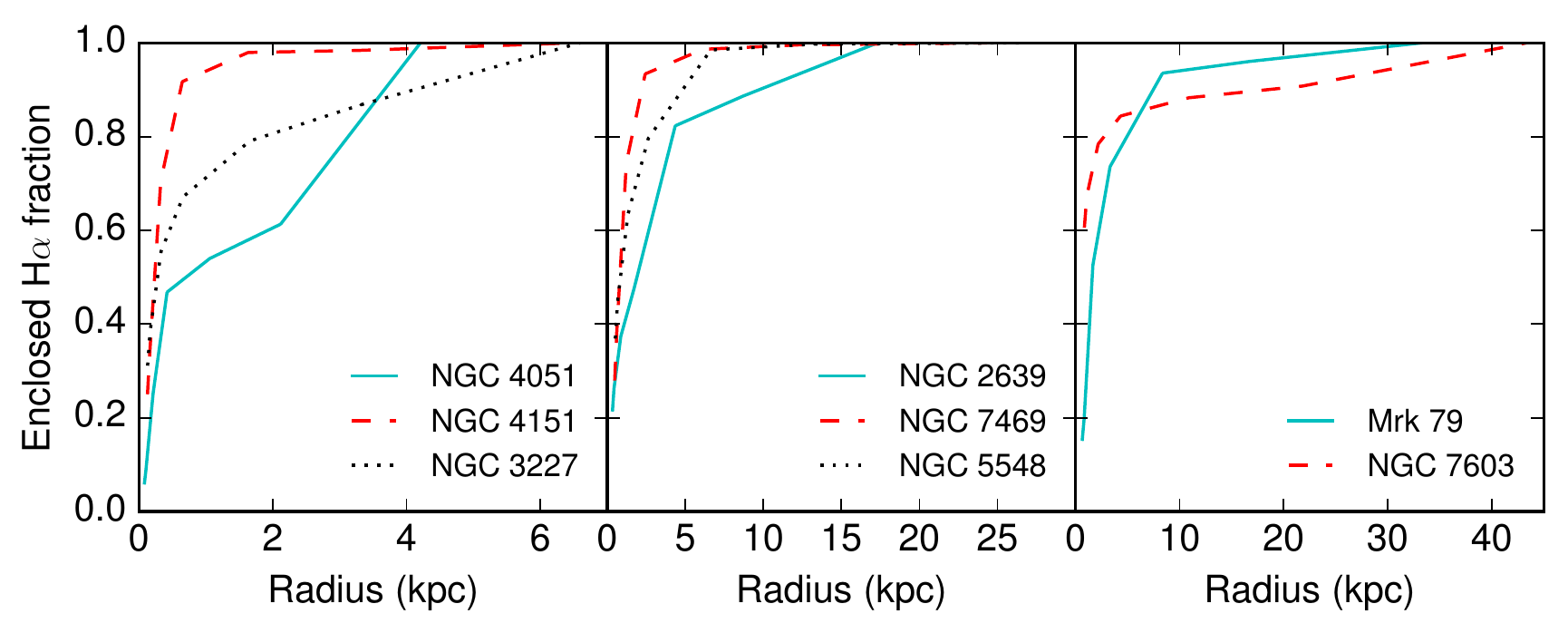}
  }
  \subfigure[Seyfert 2 and LINER]{
    \includegraphics[width=0.85\linewidth]{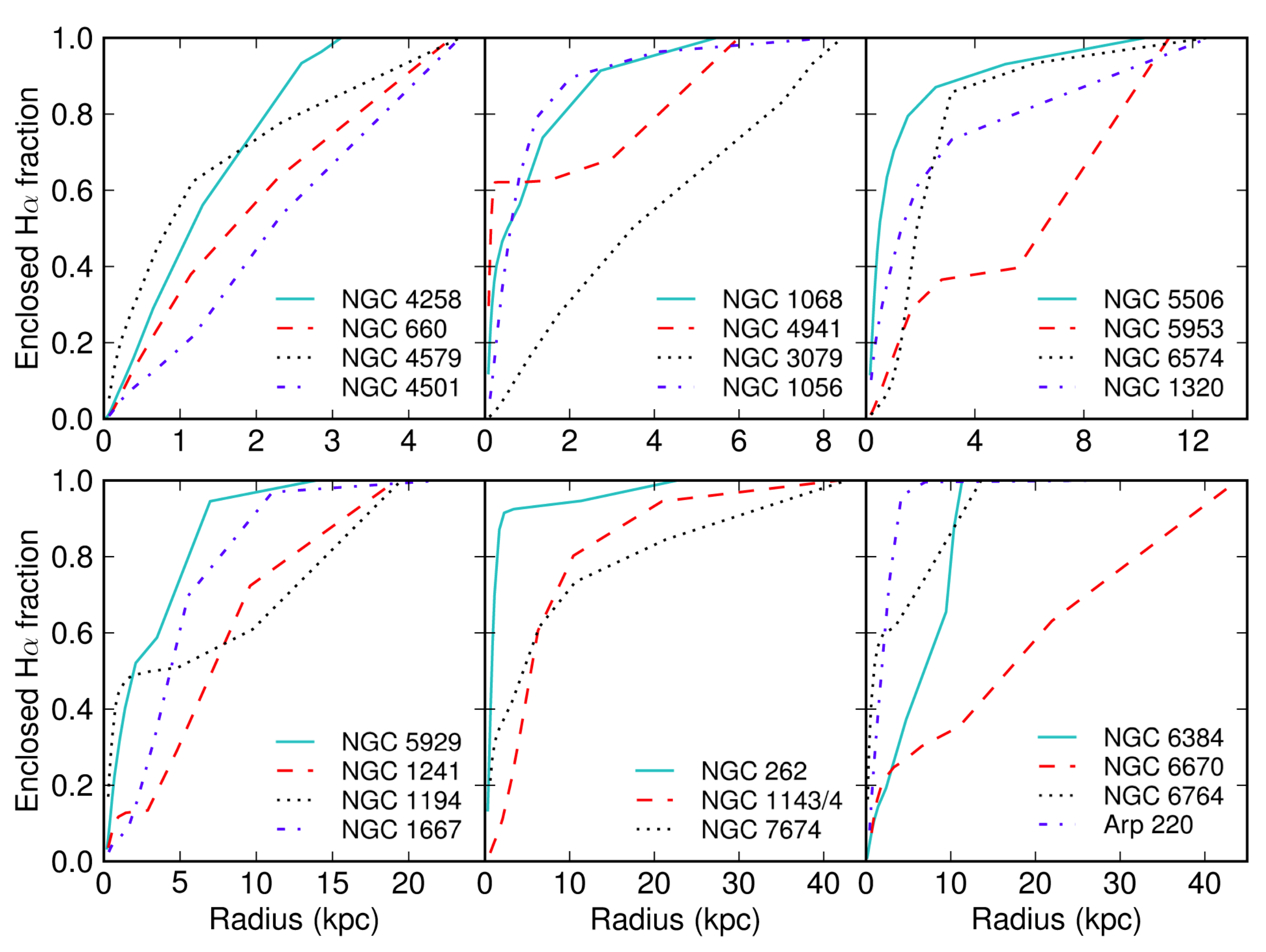}
  }
  \caption{Growth curves for Seyfert 1 and Seyfert 2 and LINER galaxies. The vertical axis represents the ratio of the \Ha\ flux within each aperture to the ``total'' flux.\label{fig:growthcurves}}
\end{minipage}
\end{figure*}

Table \ref{tab:flux} gives the integrated \Ha\ + [N~\textsc{ii}] flux of each galaxy in several selected apertures. The total flux from the AGN and host galaxy was measured with an aperture 147\arcsec\ in diameter, or 177\arcsec\ for the most extended galaxies. Figure \ref{fig:growthcurves} shows the ratio of the \Ha\ flux within each radius to the ``total'' \Ha\ flux for each galaxy. The half-light diameter of the line emission has a median value slightly larger than 15\arcsec\ in our galaxy sample. The slope of the ``growth'' curves shown in Figure \ref{fig:growthcurves} depends on both the strength of the Seyfert nucleus and the extendedness of the host galaxy; the galaxies with brighter Seyfert 1 nuclei, such as NGC 4151, tend to be more concentrated.

We defined the flux from the host galaxy alone as the total flux minus the corrected 2\farcs 9 nuclear contribution. This ``extended'' line flux is assumed to be produced in H~\textsc{ii} regions, photoionized by massive young stars. We follow previous researchers in taking the \Ha\ luminosity of the extended regions as a tracer of the ionizing photon luminosity, which should be directly proportional to the total rate of recent star formation in the galaxy.  As noted above, in a few of the least luminous AGN, this might be a small underestimate, since it could have missed an additional contribution from H~\textsc{ii} regions inside the inner 2\farcs 9, i.e. within 200 parsecs of the nucleus at the median redshift of our sample.

We corrected the observed given in Table \ref{tab:flux} for extinction by assuming one magnitude of extinction at the wavelength of \Ha. We corrected for [N~\textsc{ii}] emission by assuming that [N~\textsc{ii}] comprised 25\% of the measured \Ha\ + [N~\textsc{ii}] flux \citep{kennicutt1983}. To convert the observed fluxes into luminosities, we used Virgo-infall-corrected distances from NED. From the extended \Ha\ luminosity, we estimated star formation rates (SFRs) using the calibration given by \citet{kennicutt2012}: $$ \mathrm{SFR}_{\mathrm{H}\alpha}~(M_{\sun}~\mathrm{yr}^{-1}) = 5.37 \times 10^{-42}~L_{\mathrm{H}\alpha}~(\mathrm{erg}~\mathrm{s}^{-1}) $$ Derived luminosities and star formation rates are given in Table \ref{tab:luminosity}.

\begin{deluxetable*}{lcccccc}
\tabletypesize{\small}
\tablewidth{0pt}
\tablecaption{Derived Parameters \label{tab:luminosity}}
\tablehead{\colhead{Galaxy} & \colhead{$\log({L_{\mathrm{H}\alpha}})$} & \colhead{$\log({L_{\mathrm{H}\alpha}})$} & \colhead{$\log({L_{\mathrm{H}\alpha}})$} & \colhead{SFR} & \colhead{sSFR} & \colhead{Type} \\ 
\colhead{} & \colhead{(Nuclear)} & \colhead{(Extended)} & \colhead{(Total)} & \colhead{($M_{\sun}~\mathrm{yr}^{-1}$)} & \colhead{($\mathrm{Gyr}^{-1}$)} & \colhead{} } 
\startdata
Mrk 79 & 41.22 & 41.80 & 41.90 & 3.38 & 0.025 & Sy 1 \\
NGC 2639 & 40.71 & 41.23 & 41.33 & 0.92 & 0.011 & Sy 1 \\
NGC 3227 & 40.90 & 41.12 & 41.32 & 0.70 & \nodata & Sy 1 \\
NGC 4051 & 39.53 & 40.76 & 40.78 & 0.31 & 0.044 & Sy 1 \\
NGC 4151 & 41.28 & 41.25 & 41.57 & 0.95 & \nodata & Sy 1 \\
NGC 5548 & 41.95 & 41.90 & 42.23 & 4.26 & 0.096 & Sy 1 \\
NGC 7603 & 42.25 & 41.82 & 42.39 & 3.55 & \nodata & Sy 1 \\
NGC 7469 & 42.11 & 41.76 & 42.27 & 3.10 & 0.011 & Sy 1 \\
NGC 4258 & 38.73 & 40.76 & 40.76 & 0.31 & \nodata & Sy 1.9 \\
NGC 4579 & 40.06 & 41.13 & 41.16 & 0.72 & \nodata & Sy 1.9 \\
NGC 5506 & 40.67 & 41.29 & 41.39 & 1.05 & \nodata & Sy 1.9 \\
NGC 262 & 40.78 & 41.31 & 41.42 & 1.10 & 0.024 & Sy 2 \\
NGC 660 & 39.10 & 40.91 & 40.91 & 0.43 & 0.024 & Sy 2 \\
NGC 1056 & 40.08 & 41.03 & 41.08 & 0.58 & 0.050 & Sy 2 \\
NGC 1068 & 41.38 & 41.95 & 42.05 & 4.75 & \nodata & Sy 2 \\
NGC 1144 & 40.99 & 42.42 & 42.44 & 14.13 & \nodata & Sy 2 \\
NGC 1194 & 39.74 & 40.09 & 40.25 & 0.07 & 0.001 & Sy 2 \\
NGC 1241 & 40.45 & 41.76 & 41.78 & 3.07 & 0.037 & Sy 2 \\
NGC 1320 & 40.47 & 41.27 & 41.34 & 1.01 & 0.050 & Sy 2 \\
NGC 1667 & 40.65 & 42.11 & 42.12 & 6.91 & 0.063 & Sy 2 \\
NGC 3079 & 39.20 & 41.30 & 41.31 & 1.08 & \nodata & Sy 2 \\
NGC 4501 & 39.82 & 41.51 & 41.52 & 1.74 & 0.050 & Sy 2 \\
NGC 4941 & 39.83 & 39.62 & 40.04 & 0.02 & 0.002 & Sy 2 \\
NGC 5929 & 40.38 & 41.58 & 41.61 & 2.05 & 0.039 & Sy 2 \\
NGC 5953 & 40.25 & 41.98 & 41.99 & 5.15 & 0.172 & Sy 2 \\
NGC 6574 & 40.12 & 41.93 & 41.94 & 4.56 & \nodata & Sy 2 \\
NGC 7674 & 41.83 & 42.42 & 42.52 & 14.13 & 0.009 & Sy 2 \\
NGC 6384 & 39.66 & 41.49 & 41.49 & 1.65 & \nodata & LINER \\
NGC 6670 & 41.14 & 42.14 & 42.18 & 7.45 & 0.047 & non-Sy \\
NGC 6764 & 41.20 & 41.49 & 41.67 & 1.64 & \nodata & LINER \\
Arp 220 & 40.47 & 41.23 & 41.30 & 0.92 & \nodata & LINER \\
\enddata
\tablecomments{~Luminosities are in units of $\mathrm{erg}~\mathrm{s}^{-1}$. Luminosities are corrected for extinction, [N~\textsc{ii}], and aperture losses. Certain galaxies are missing sSFRs because either IRAC images were unavailable, or the IRAC images were saturated (see Table \ref{tab:comparisons}).}
\end{deluxetable*}

The extinction and [N~\textsc{ii}] corrections we used were those applied by \citet{kennicutt1998} to derive their \Lha-SFR relation; however, extinction varies between individual galaxies in the sample, introducing some scatter into this relation. Although for most of the galaxies in our sample, nuclear extinction can be estimated from Balmer emission line ratios \citep[Malkan et al.\ 2016, in preparation; ][]{moustakas2006}, only a small number of \emph{integrated} Balmer decrements are available in order to estimate extinction in the bodies of the host galaxies. The measurements of \citet{moustakas2006} suggest that our assumed 1 magnitude correction might be $\sim 30\%$ too high for Seyfert 1s, and $\sim 30 \%$ too low for Seyfert 2s. These discrepancies are not much larger than other uncertainties, and are based on only 4 Seyfert 1s.  Below we find no clear evidence that we have underestimated the SFR in Seyfert 2s compared with Seyfert 1s, so we elect to apply the same extinction correction to each type.

\section{Results}

\subsection{Star Formation Rate Comparisons}

\begin{deluxetable*}{lcccccc}
\tabletypesize{\small}
\tablewidth{0pt}
\tablecaption{Supplementary Data \label{tab:comparisons}}
\tablehead{\colhead{Galaxy} & \colhead{$\log(L_{\mathrm{X}})$\tablenotemark{a}} & \colhead{$\log(L_{7.7\micron})$\tablenotemark{b}} & \colhead{$\log(L_{\mathrm{FIR}})$\tablenotemark{c}} & \colhead{$\log(L_{1.4\mathrm{GHz}})$\tablenotemark{d}} & \colhead{Type} & \colhead{Reference} \\ 
\colhead{} & \colhead{($\mathrm{erg}~\mathrm{s}^{-1}$)} & \colhead{($\mathrm{erg}~\mathrm{s}^{-1}$)} & \colhead{($\mathrm{erg}~\mathrm{s}^{-1}$)} & \colhead{($\mathrm{erg}~\mathrm{s}^{-1}$)} & \colhead{} & \colhead{} } 
\startdata
Mrk 79 & 43.78 & 43.52 & 44.21 & 38.44 & Sy 1 & 1 \\
NGC 2639 & 40.14 & 42.93 & \nodata & 38.66 & Sy 1 & 2 \\
NGC 3227 & 41.57 & \nodata & 42.98 & 37.33 & Sy 1 & 1 \\
NGC 4051 & 40.91 & 42.58 & 42.69 & 36.53 & Sy 1 & 1 \\
NGC 4151 & 42.09 & \nodata & 43.05 & 37.77 & Sy 1 & 1 \\
NGC 5548 & 43.43 & 43.58 & 43.90 & 38.46 & Sy 1 & 1 \\
NGC 7603 & \nodata & \nodata & 44.31 & 38.90 & Sy 1 & 1 \\
NGC 7469 & 43.25 & 43.91 & 45.07 & 39.19 & Sy 1 & 1 \\
NGC 4258 & \nodata & \nodata & \nodata & 37.85 & Sy 1.9 & 2 \\
NGC 4579 & 41.36 & \nodata & 43.40 & 37.45 & Sy 1.9 & 2 \\
NGC 5506 & 42.85 & \nodata & 43.58 & 38.26 & Sy 1.9 & 1 \\
NGC 262 & 43.34 & 43.19 & 44.02 & 39.30 & Sy 2 & 2 \\
NGC 660 & 39.41 & 43.16 & 44.08 & 38.01 & Sy 2 & 2 \\
NGC 1056 & \nodata & 42.78 & \nodata & 37.51 & Sy 2 & 3 \\
NGC 1068 & 42.15 & \nodata & 44.70 & 39.21 & Sy 2 & 1 \\
NGC 1144 & 43.61 & \nodata & 44.74 & 39.55 & Sy 2 & 1 \\
NGC 1194 & 42.32 & 42.99 & 43.73 & 37.10 & Sy 2 & 3 \\
NGC 1241 & \nodata & 43.54 & 43.56 & \nodata & Sy 2 & \nodata \\
NGC 1320 & 42.65 & 42.73 & 43.67 & 37.13 & Sy 2 & 2 \\
NGC 1667 & 42.76 & 43.86 & 44.29 & 38.67 & Sy 2 & 2 \\
NGC 3079 & 40.87 & \nodata & 43.79 & 38.70 & Sy 2 & 2 \\
NGC 4501 & 41.69 & 42.98 & 43.88 & 37.90 & Sy 2 & 3 \\
NGC 4941 & \nodata & 42.03 & \nodata & 36.98 & Sy 2 & 2 \\
NGC 5929 & \nodata & 43.25 & 44.05 & 38.44 & Sy 2 & 1 \\
NGC 5953 & \nodata & 43.40 & 43.75 & 38.05 & Sy 2 & 1 \\
NGC 6574 & \nodata & \nodata & \nodata & 38.32 & Sy 2 & 2 \\
NGC 7674 & 43.62 & 43.90 & 44.94 & 39.77 & Sy 2 & 1 \\
NGC 6384 & \nodata & \nodata & \nodata & 37.31 & LINER & 3 \\
NGC 6670 & \nodata & 44.34 & \nodata & 39.19 & non-Sy & 2 \\
NGC 6764 & \nodata & \nodata & \nodata & 38.43 & LINER & 2 \\
Arp 220 & \nodata & \nodata & 45.669\tablenotemark{e} & 39.48 & LINER & 1 \\
\enddata
\tablenotetext{a}{From \citet{brightman2011}.}
\tablenotetext{b}{Measured in this paper from reduced Spitzer images (program ID 3269, PI: Gallimore).}
\tablenotetext{c}{From \citet{rush1996}, unless otherwise specified.}
\tablenotetext{d}{References refer to $1.4~\mathrm{GHz}$ luminosities.}
\tablenotetext{e}{From \citet{spinoglio2002}.}
\tablerefs{(1) \citet{rush1996}; (2) \citet{condon1998}; (3) \citet{condon2002}}
\end{deluxetable*}

\begin{figure}[htb]
  \centering
  \includegraphics[width=\linewidth]{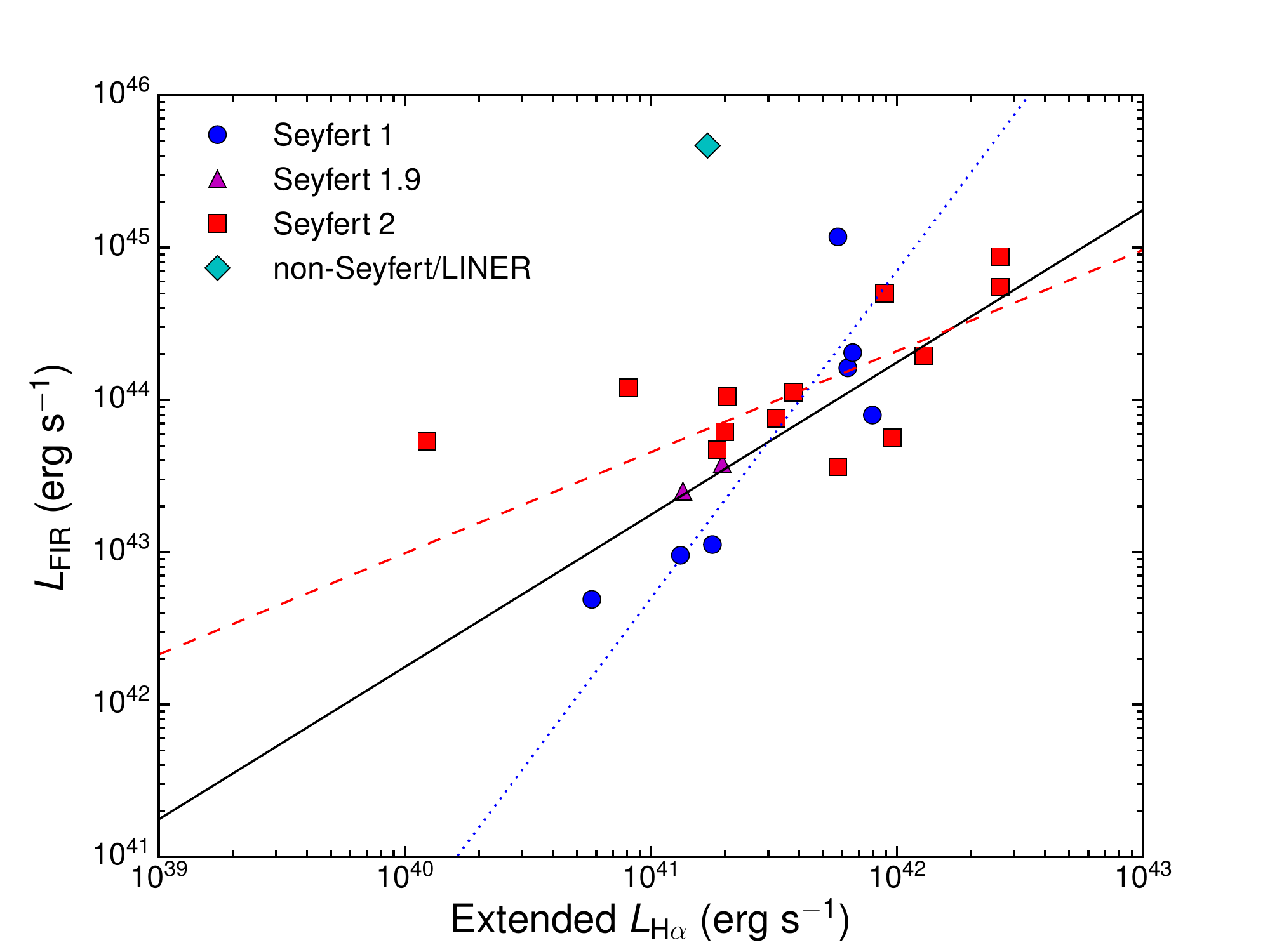}
  \caption{Extended \Ha\ versus FIR luminosity. Dotted and dashed lines represent linear least squares fits of Seyfert 1s and Seyfert 2s. The outlying Sy 2 with very weak extended \Ha\ emission is NGC 1194, and the LINER with exceptionally strong FIR emission is Arp 220. The solid line represents equivalence between the \Ha\ and FIR SFR calibrations. Most of the galaxies in our sample lie roughly along this line, indicating agreement between the two SFR relations, although there is substantial scatter.\label{fig:FIR}}
\end{figure}

We compared star formation rates derived from our \Ha\ measurements with other star formation rate estimators proposed in the literature: far-infrared luminosity, the $7.7~\micron$ PAH emission feature, and 1.4~GHz luminosity. Table \ref{tab:comparisons} summarizes the values we used for these comparisons. In Figure \ref{fig:FIR} we compare extended \Lha\ to far-infrared luminosity \citep[from][]{spinoglio1995}. Linear least squares fits of the log-log plot gave a slope of 2.15 for Seyfert 1 galaxies (blue dotted line) and 0.66 for Seyfert 1.9 and 2 galaxies (red dashed line), with RMS scatters about the best-fit line of 0.45 dex and 0.16 dex for Seyfert 1s and 2s respectively. We estimated SFRs from the far-IR luminosities with the calibration of \citet{kennicutt1998}: $$ \mathrm{SFR}_{\mathrm{FIR}}~(M_{\sun}~\mathrm{yr}^{-1}) = 4.5 \times 10^{-44}~L_{\mathrm{FIR}}~(\mathrm{erg}~\mathrm{s}^{-1}) $$

We then set this relation equal to the \Ha\ SFR calibration of \citet{kennicutt1998} to obtain a predicted linear relationship between $L_{\mathrm{FIR}}$ and \Lha, shown as a solid line on Figure 5 with a slope of 1, not fitted to the data.\footnote{We used the \Ha\ SFR calibration of \citet{kennicutt1998} in our comparisons with the far-IR SFR, as both relations assume a \citet{salpeter} IMF, whereas the SFR calibrations given in \citet{kennicutt2012} assume a \citet{kroupa2003} IMF.} There is substantial scatter between these two estimators of star formation in the host galaxies, although there is no significant offset between the estimates from the far-IR and \Ha\ luminosity, for either Seyfert type.  A few galaxies have a much higher FIR luminosity than their \Ha\ luminosities would suggest.  They could have extra far-IR emission originating not in H~\textsc{ii} regions, but in cold ``cirrus'' dust, which is illuminated by older stars.

This refers to the total far-IR luminosity without the nuclear contribution subtracted. However, \citet{spinoglio2002} argued that the AGN usually makes a relatively small contribution to the far-IR luminosity. We note that we have taken the far-IR luminosities from \citet[their Table 3]{spinoglio1995}. These are integrated across the four IRAS bands, from 12 to 100 \micron, because this is also how far-IR luminosities have generally been measured in studies which attempted to correlate these with star formation rates. If we had instead included the colder dust contribution measured out to 200 \micron\ by ISOPHOT \citep[from][]{spinoglio2002}, this would merely increase all the far-IR luminosities by a nearly constant 40\%. That adjustment would not significantly alter our conclusions.

\begin{figure}[htb]
  \centering
  \includegraphics[width=\linewidth]{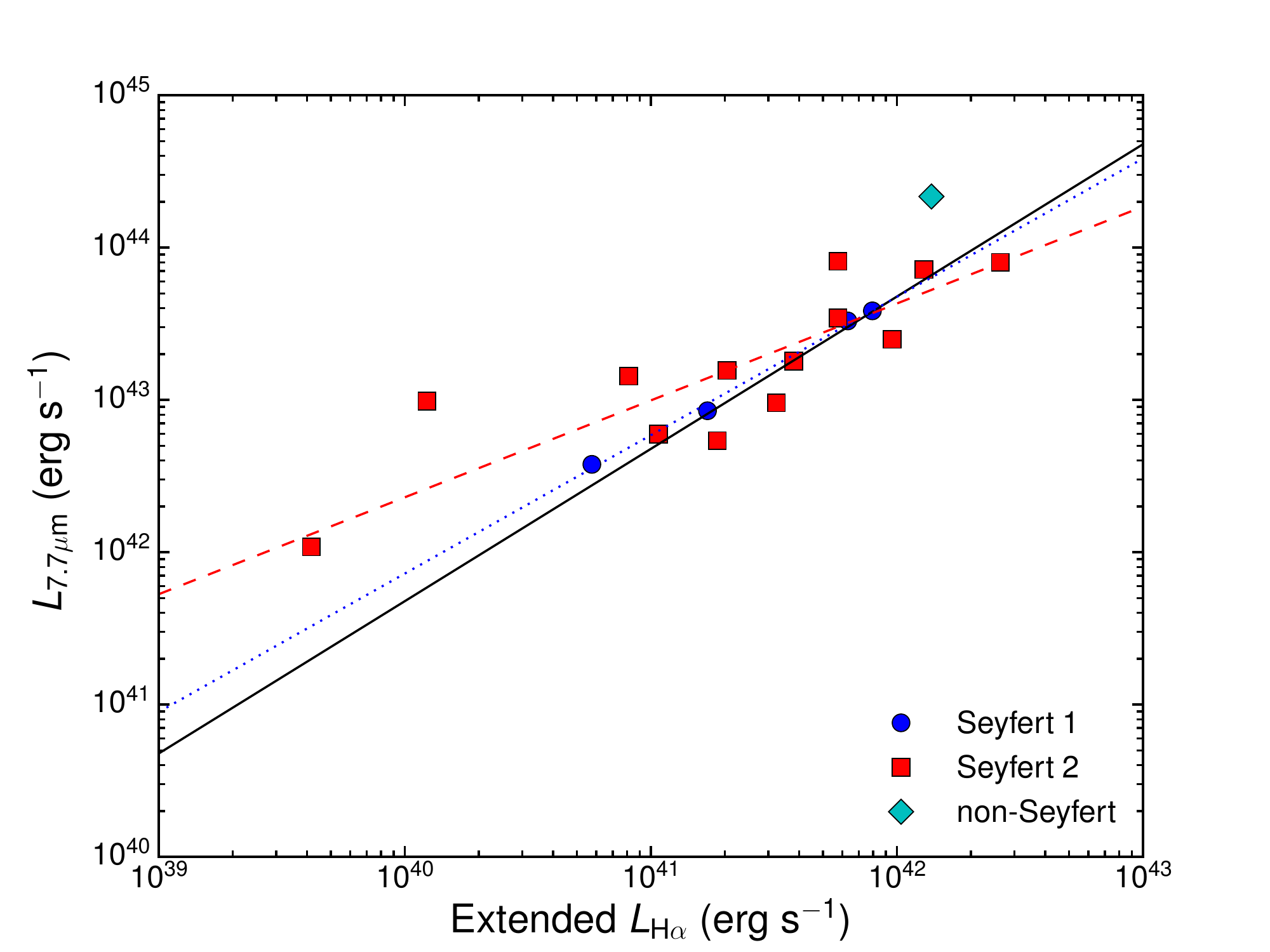}
  \caption{Extended \Ha\ versus 7.7~\micron\ PAH luminosity. Dotted and dashed lines represent linear least squares fits of Seyfert 1s and Seyfert 2s respectively. The outlying Sy 2 with very weak \Ha\ emission is NGC 1194. The solid line represents equivalence between the \Ha\ and 7.7~\micron\ PAH SFR calibrations. These two relations show rough agreement, although there is substantial scatter.}\label{fig:PAH}
\end{figure}

Next we compared the SFR derived from extended \Ha\ luminosity to that from the 7.7~\micron\ PAH emission feature. To measure this feature, we used reduced Spitzer images (program ID 3269, PI: Gallimore) in IRAC Bands 1 and 4 (3.6 and 8.0~\micron\ respectively). We measured fluxes in each band using circular aperture photometry over the same apertures used for our \Ha\ measurements, and with the flux calibration given in the IRAC instrument handbook\footnote{\url{http://irsa.ipac.caltech.edu/data/SPITZER/docs/irac/iracinstrumenthandbook/}}. We assumed that Band 1 contained purely starlight, dominated by red giants, and that Band 4 contained only PAH emission plus the Rayleigh-Jeans tail of the red giant starlight, without any contribution from the hot dust continuum \citep[see][]{meidt2012}. To isolate the PAH feature, we multiplied the Band 1 flux by a Rayleigh-Jeans $\nu^2$ factor and subtracted this quantity from the band 4 flux\footnote{We chose the IRAC band 1 images to estimate the starlight because they have the best SNR and sharpest PSF to separate the nucleus from the surrounding host galaxy. We also checked our starlight extrapolation to longer wavelengths by measuring the starlight in IRAC Band 2 (4.5~\micron).  In most cases, the annular flux we measured in Band 2 is reasonably consistent with that measured in Band 1, for stars following a roughly Rayleigh-Jeans fall-off: $F(4.5~\micron) = 0.7 F(3.6~\micron)$. The 4.5~\micron\ annular fluxes of some Seyfert 1 galaxies in Band 2 are about 30\% higher than this. We suspect that the starlight estimates extrapolated from 3.6~\micron\ are more accurate, since some fraction of the strong 4.5~\micron\ nonstellar AGN continuum in those Seyfert 1s may have spilled out into our measuring annulus. But even if we had relied on the observed 4.5~\micron\ annular fluxes, they would not change our PAH flux estimates much, as the Band 4 (8.0~\micron) flux would still be strongly dominated by PAHs, not starlight.}. We converted these fluxes to luminosities using the Virgo-infall-corrected distance from NED. As with the \Ha\ measurements, we subtracted the central 2\farcs 9 contribution. Figure \ref{fig:PAH} compares extended \Lha\ with $L_{7.7~\mu\mathrm{m}}$. A linear least squares fit of the log-log plot gave slopes of 0.91 and 0.64 for Seyfert 1s and 2s respectively, with RMS scatters about the best-fit line of 0.05 and 0.11 dex. We estimated SFRs from the 7.7~\micron\ PAH feature with the calibration from \citet{wu2005}: $$ \mathrm{SFR}_{8~\mu\mathrm{m}} (M_{\sun}~\mathrm{yr}^{-1}) = \frac{\nu L_{\nu} (8~\micron)}{1.57 \times 10^9~L_{\sun}} $$

We compared these results with the \Ha\ SFRs obtained from our data and the relation from \citet{kennicutt1998}\footnote{As with the far-IR comparisons, we used the \Ha\ SFR calibration of \citet{kennicutt1998} here because the PAH SFR calibration given in \citet{wu2005} also assumes a Salpeter IMF.}. The solid line in Figure \ref{fig:PAH} represents a 1:1 correlation between these two relations. As with the far-IR SFR estimator, the PAH estimator yields SFRs that are on average consistent with our \Ha\ luminosities, with some scatter. Some results \citep[\eg][]{smith2007,dsr2010} suggest that the strength of the 7.7~\micron\ PAH feature is suppressed in local Seyfert nuclei. In this paper, however, only off-nuclear regions are used to measure PAH emission.

\begin{figure}[thb]
  \centering
  \includegraphics[width=\linewidth]{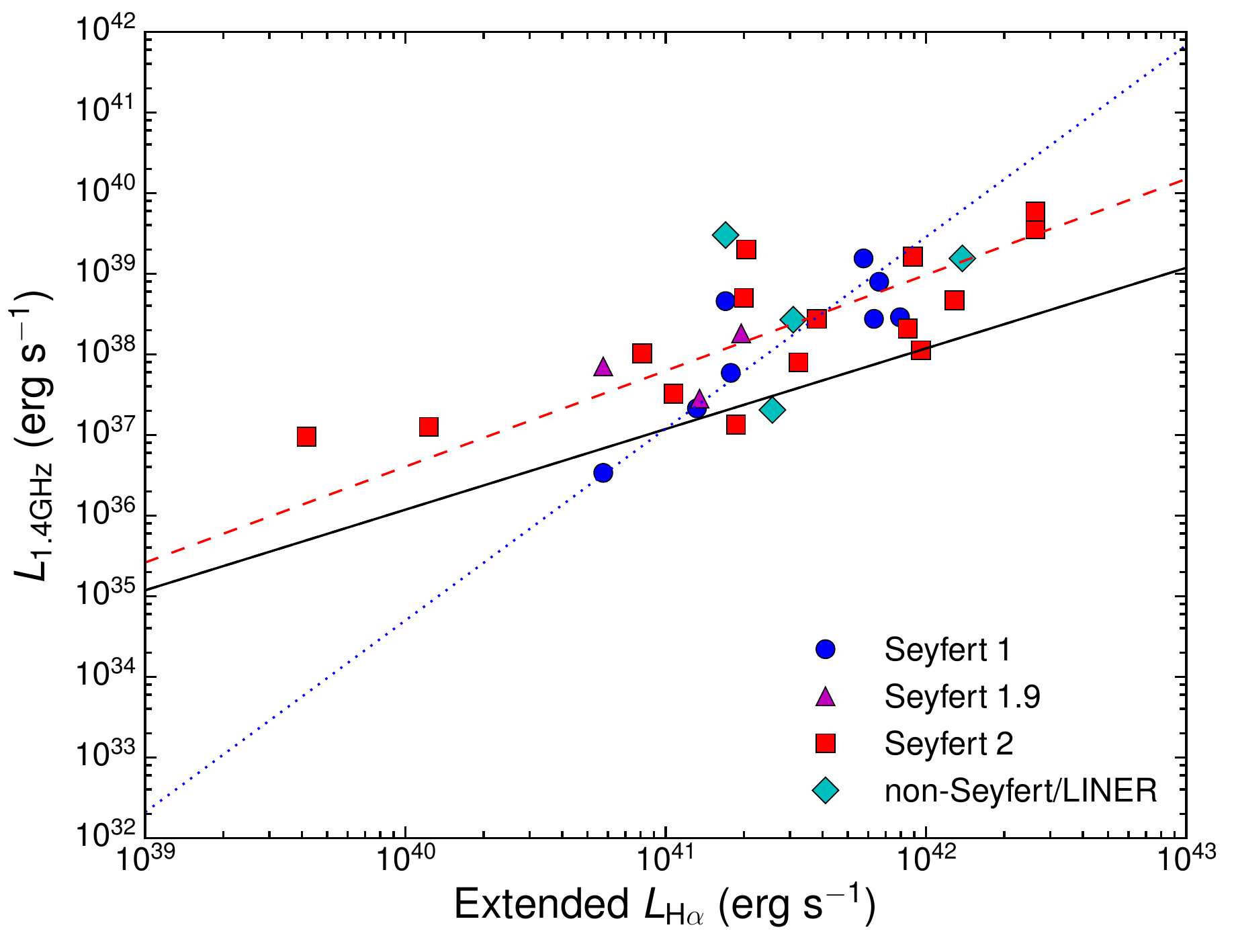}
  \caption{Extended \Ha\ versus 1.4~GHz luminosity. Dotted and dashed lines represent linear least squares fits of Seyfert 1s and Seyfert 2s respectively. The solid line represents equivalence between the \Ha\ and 1.4~GHz SFR calibrations. Many of the galaxies in our sample lie above this line due to the presence of the Seyfert nucleus.\label{fig:radio}}
\end{figure}

Figure \ref{fig:radio} compares \Lha\ with 1.4~GHz luminosity from \citet{rush1996} and the NRAO VLA Sky Survey \citep[NVSS; ][]{condon1998}. A linear least-squares fit of the log-log plot gave slopes of 2.38 for Seyfert 1s and 1.19 for Seyfert 2s, with RMS scatters about the best-fit line of 0.47 and 0.18 dex. We estimated SFRs from $L_{1.4~\mathrm{GHz}}$ with the calibration of \citet{kennicutt2012}: $$ \mathrm{SFR}_{1.4~\mathrm{GHz}} (M_{\sun}~\mathrm{yr}^{-1}) = 6.35\times10^{-29}~L_{1.4~\mathrm{GHz}}~(\mathrm{erg}~\mathrm{s}^{-1}~\mathrm{Hz}^{-1}) $$

The solid line in Figure \ref{fig:radio} represents equality between the 1.4~GHz and \Ha\ SFR relations. The Seyfert galaxies in our sample lie on average 0.8 dex above this line, indicating a higher radio luminosity than can be attributed to star formation alone. This is presumably due to the presence of the Seyfert nucleus, although it is also possible that the integrated radio luminosities include a contribution from star formation in and around the nucleus. \citet{rush1996} drew a similar conclusion about the AGN contribution to the observed radio luminosities of the 12~\micron\ Seyfert galaxies.

\subsection{Specific Star Formation Rates}

To estimate specific star formation rates (sSFRs, defined as $\mathrm{sSFR} = \mathrm{SFR}/M_*$) for our galaxy sample, we estimated stellar masses from the Band 1 3.6~\micron\ extended luminosities (with ``nuclear'' contribution removed) by assuming a 3.6~\micron\ mass-to-light ratio of 9.77 \citep{zhu2010}. We then measured sSFRs for each galaxy, given in Table \ref{tab:luminosity}, using our \Ha\ SFRs. The specific star formation rates of most of our Seyfert host galaxies are $0.05~\mathrm{Gyr}^{-1} \pm 50\%$, consistent with normal star-forming spirals in the local Universe \citep{brinchmann2004}.

\subsection{AGN and Host Galaxy Star Formation}

\begin{figure}[thb]
  \centering
  \includegraphics[width=\linewidth]{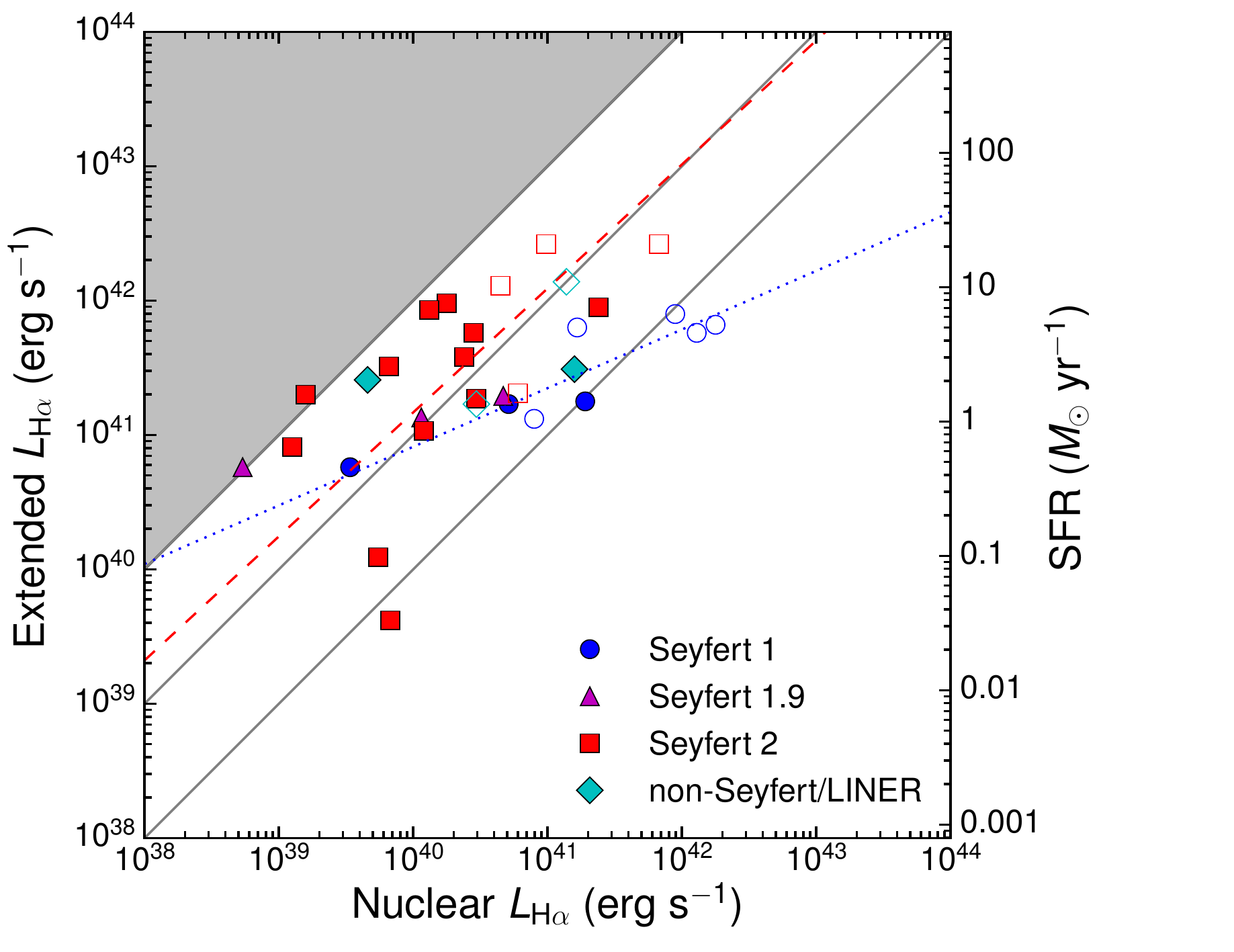}
  \caption{``Nuclear'' versus extended \Ha\ luminosity. SFRs are given on the right-hand axis. Dotted and dashed lines represent linear least squares fits of Seyfert 1s and Seyfert 2s respectively. Open symbols denote galaxies with $z > 0.015$. The solid lines represent nuclear-to-extended ratios of 0.01, 0.1, and 1. Galaxies with a nuclear luminosity less than 1\% of the extended would be misclassified as normal star-forming galaxies and are thus very likely to have been excluded from this plot.\label{fig:agn}}
\end{figure}

Figure \ref{fig:agn} compares the ``nuclear'' \Ha\ luminosity to that from the host galaxy. This is only a rough attempt to measure nuclear emission, since some of this nuclear luminosity originates in H~\textsc{ii} regions near the galaxy center. Nonetheless, these quantities appear to be correlated. A proper linear least squares fit of the log-log plot gave slopes of 0.44 and 0.92 for Seyfert 1s and 2s respectively, with RMS scatters about the best-fit line of 0.08 and 0.18 dex. The solid lines show a nuclear luminosity equal to 1\%, 10\%, and 100\% of the extended. This correlation might be due in part to selection effects; since the sample is flux-limited, the highest-luminosity galaxies are also the highest in redshift. Galaxies with low AGN luminosity and high extended luminosity would have been excluded from the original sample, and thus cannot appear on this plot. This effect appears to set in when the nuclear luminosity is below 1\% of the extended \Ha\ luminosity. However, selection effects cannot account for the absence of galaxies with a high AGN luminosity and a low star-formation rate. This argues that the apparent correlation is real, although the scatter is large. The Seyfert 1s generally have a higher nuclear-to-extended ratio; some have a nuclear luminosity---which includes a broad line region---comparable to their extended luminosity.

An inherent uncertainty in this method lies in the fact that although we define a 2\farcs 9 aperture as the ``nucleus'' of the galaxy, the corresponding physical area varies by an order of magnitude between galaxies in the sample. At larger redshifts, star formation within this aperture may be contributing to the so-called ``nuclear'' \Lha, which we attributed to the AGN. In Figure \ref{fig:agn}, the galaxies with $z > 0.015$ are denoted by open symbols. If our measure of the AGN luminosity for these higher-redshift galaxies were systematically contaminated by star formation within 0.5 kpc of the galactic center, the extended \Lha\ would be correspondingly lower. Thus the highest-redshift galaxies would be shifted to the right of Figure \ref{fig:agn}. However, the galaxies with $z > 0.015$ show the same correlation as the lower-redshift galaxies, suggesting that contamination from circumnuclear star formation is not a significant effect. Even if the AGN luminosity we calculated is only an upper limit at high redshift, our conclusions are not significantly changed. 

\citet{dsr2012} found a similar AGN/star formation correlation in nearby Seyfert galaxies, although they used mid-IR emission features, rather than \Ha, to estimate both quantities. They found that the correlation of AGN to young-star luminosities improved when they considered only the minority of star formation within 1 kpc of the galaxy centers, in effect the ``circumnuclear'' star formation. To examine this, we measured the SFR in an annulus between 2\farcs 9 and 1 kpc for the subset of galaxies in our sample for which the inner kpc subtends less than 2\farcs 9. We found that considering star formation only within the inner kpc of the host galaxy did not tighten the correlation between SFR and nuclear \Ha\ luminosity. However, this technique is not sensitive to star formation within 1 kpc of the galactic center for $ \sim 1/4$ of the galaxies in our sample, or within 0.5 kpc for $\sim 1/2$ of the galaxies in our sample. Thus, although our imaging provided no evidence that the SFR correlations we presented would have changed if we could have isolated the inner kpc of the host galaxy, our ground-based, seeing-limited study is not suited to address this question.

\subsection{AGN \Ha\ and Hard X-Ray Luminosity}

\begin{figure}[thb]
  \centering
  \includegraphics[width=\linewidth]{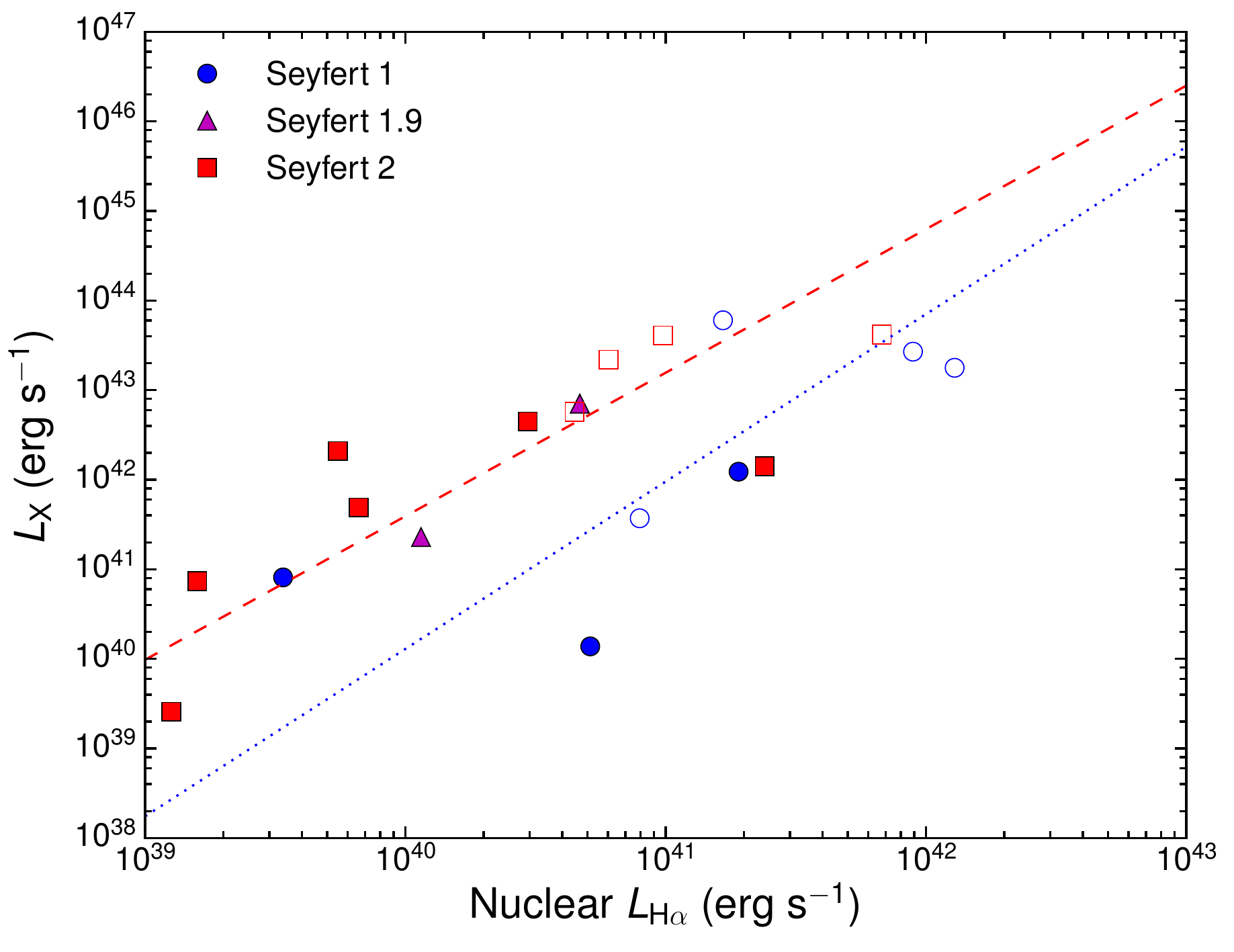}
  \caption{Nuclear \Ha\ vs hard X-ray luminosity. Open symbols denote galaxies with $z > 0.015$. Dotted and dashed lines represent linear least squares fits of Seyfert 1s and 2s respectively. Seyfert 1s are offset from Seyfert 2s by 1 dex on average due to the presence of broad-line regions.\label{fig:xray}}
\end{figure}

Figure \ref{fig:xray} compares nuclear \Lha\ with hard X-ray luminosity \citep{brightman2011}. Seyfert 1s show greater nuclear \Ha\ luminosities than Seyfert 2s by 1 dex on average, due to the presence of broad-line regions. In other words, the broad-line contribution to the nuclear \Ha\ luminosity in Seyfert 1s is roughly 10 times stronger on average than the narrow-line component. This offset is also present in the nuclear spectroscopic \Ha\ fluxes from Malkan et al. (2016, in preparation) shown in Figure \ref{fig:spec}. A linear least-squares fit of the log-log plot gives slopes of 1.87 and 1.60 for Seyfert 1s and 2s respectively, with RMS scatters about the best-fit line of 0.49 and 0.27 dex. Due to these high scatters, it would be problematic to predict either nuclear \Lha\ or $L_\mathrm{X}$ based on the other.

\begin{figure}[thb]
  \centering
  \includegraphics[width=\linewidth]{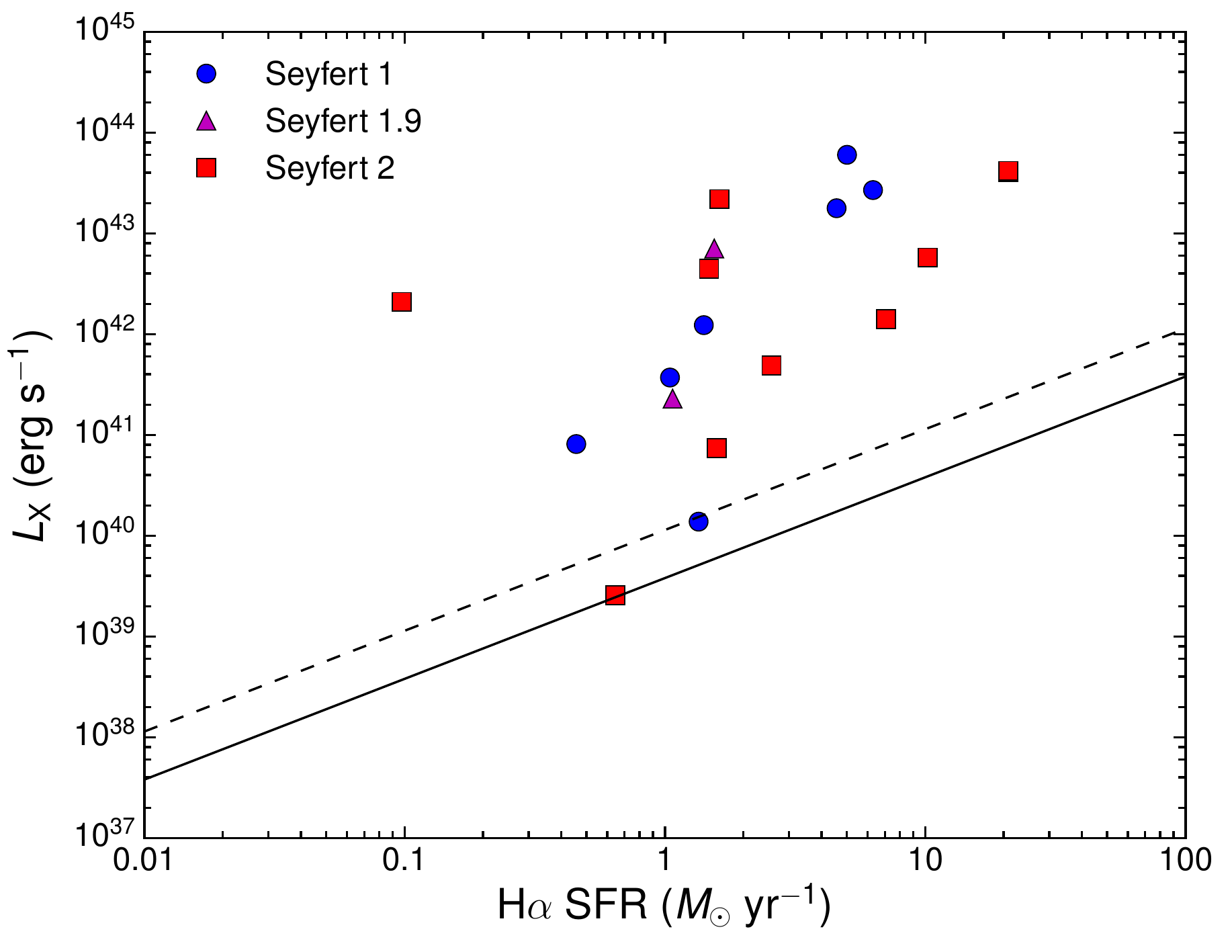}
  \caption{\Ha\ SFR vs X-ray luminosity. The solid line represents the $L_{\mathrm{X}}$-SFR relation from \citet{persic2007}. The dashed line represents $L_{\mathrm{X}}$ three times that predicted by this relation. Galaxies above this line are classified as AGN candidates by \citet{lehmer2008}.\label{fig:sfrlx}}
\end{figure}

Figure \ref{fig:sfrlx} compares $L_\mathrm{X}$ with the \Ha\ SFR, as in \citet{lehmer2008}. The solid line represents the $L_\mathrm{X}$-SFR relation from \citet{persic2007}. \citet{lehmer2008} classified galaxies with $L_\mathrm{X}$ more than three times the value predicted by this relation (shown as a dashed line in Figure \ref{fig:sfrlx}) as AGN candidates. All of the galaxies in our sample (with the exception of NGC 660) lie above this line. However, one of our weakest AGN---NGC 2639---only barely meets this criterion.

\subsection{Misclassification of Seyfert Galaxies at Large Distances}

\begin{figure*}[htb]
\begin{minipage}{180mm}
  \centering
  \subfigure[Seyfert 1]{
    \includegraphics[width=0.85\linewidth]{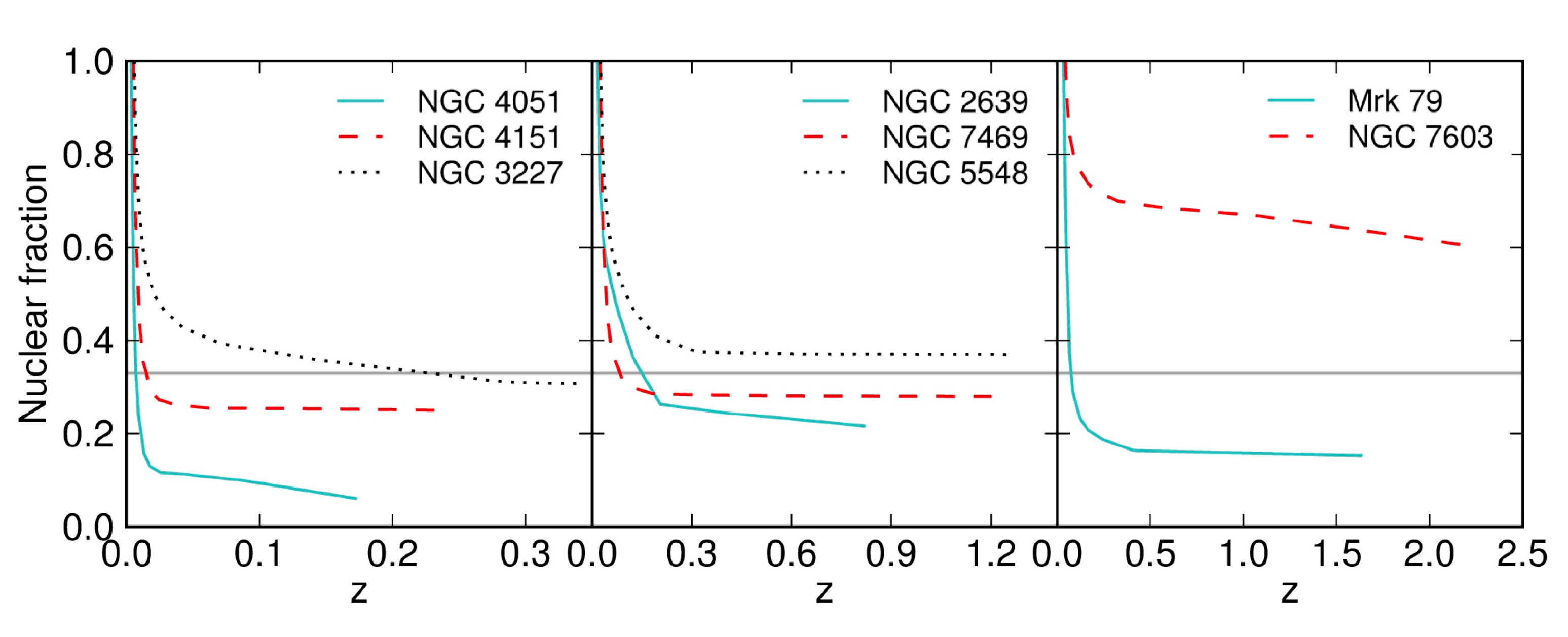}
  }
  \subfigure[Seyfert 2 and LINER]{
    \includegraphics[width=0.85\linewidth]{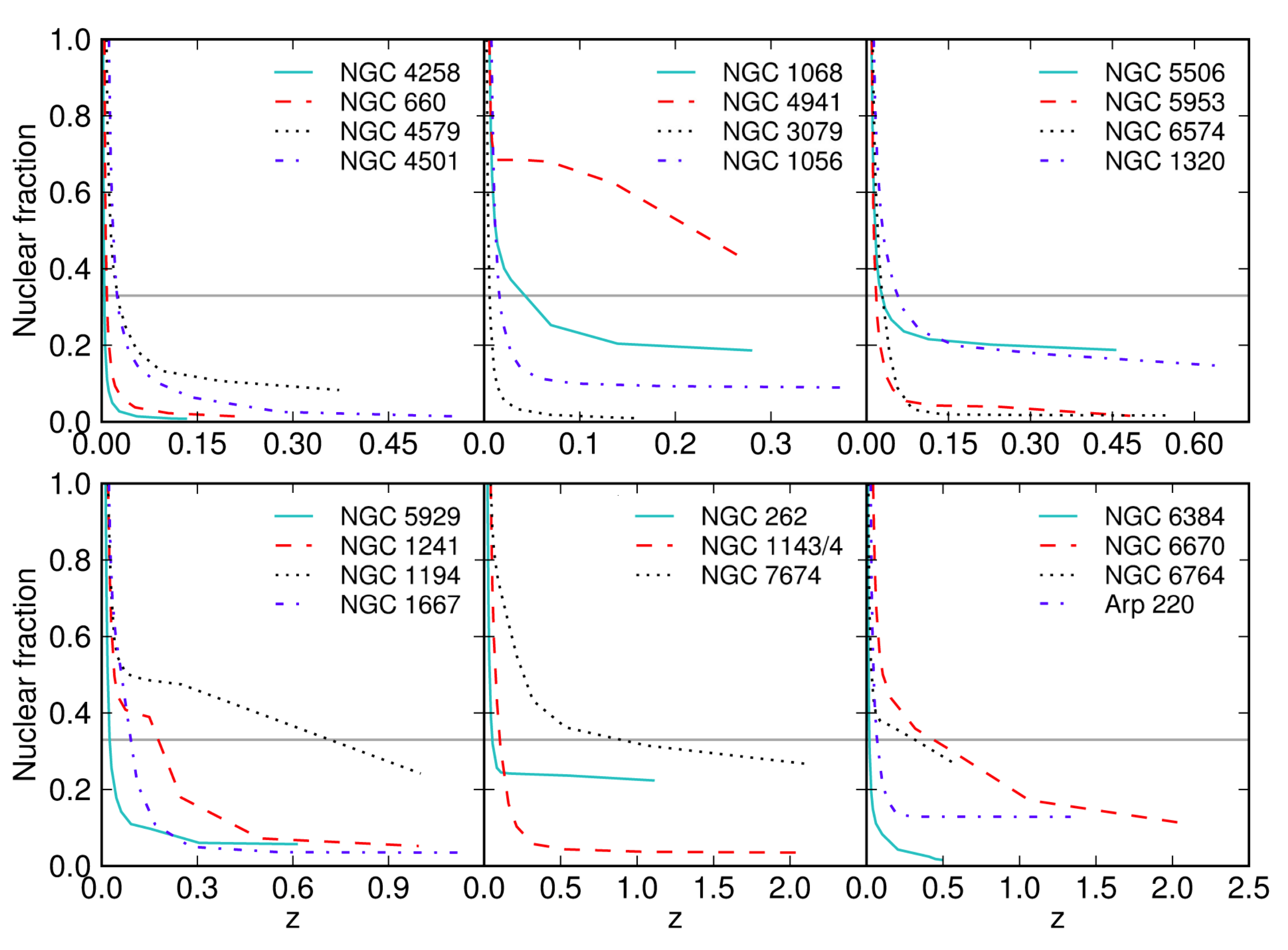}
  }
  \caption{Nuclear fraction versus redshift for Seyfert 1 and Seyfert 2 and LINER galaxies. The nuclear fraction is the ratio of the ``nuclear'' \Ha\ flux to that enclosed within each physical radius. The horizontal axis gives the redshift at which this physical radius corresponds to a fixed 2\arcsec\ aperture diameter. The redshift at which these curves drop below a nuclear fraction of $1/3$ (shown as a dashed line) represents the redshift at which these Seyferts would be misclassified as normal star-forming galaxies.\label{fig:nuclearfraction}}
\end{minipage}
\end{figure*}

From our measured equivalent widths, we determined that most of the galaxies in this sample would satisfy equivalent width limits of typical high-redshift spectroscopic surveys. Based on our plots of \Ha\ equivalent width versus radius from the galaxy center in Figure 2, only six of the galaxies in our sample have equivalent widths that drop below 20~\AA\ in the largest aperture. However, whether the galaxy is identified as a Seyfert depends on the origin of these emission lines. If a large enough fraction of the \Ha\ emission originates in H~\textsc{ii} regions in the host galaxy, the galaxy's location on line ratio diagrams such as [N~\textsc{ii}]/\Ha\ vs. [O~\textsc{iii}]/H$\beta$ \citep{bpt} shifts into the realm of normal galaxies. This may explain the incompleteness of Seyfert spectroscopic surveys beyond the local universe: higher-redshift Seyferts may often be misclassified as normal star-forming galaxies, due to contamination from star formation in the host galaxy.

Figure \ref{fig:nuclearfraction} simulates how each of the galaxies in our sample would appear in an SDSS fiber spectrum at a range of redshifts. As in Figure 4, we converted each aperture over which we integrated the flux into a physical radius (kpc) using the galaxy's Virgo-infall-corrected distance from NED. The quantity plotted on the vertical axis of Figure \ref{fig:nuclearfraction} is the nuclear fraction, defined as the ratio of the ``nuclear'' \Ha\ flux to the total \Ha\ flux enclosed within each radius. The horizontal axis represents the redshift at which each physical radius would correspond to a fixed angular diameter of 2\arcsec. We then interpolated over each curve to determine the redshift at which the nuclear fraction drops below $1/3$; i.e., the redshift at which only one third of the \Ha\ emission visible in a 2\arcsec\ aperture originates in the galaxy center. We selected $1/3$ as a typical threshold below which most observations would have difficulty identifying a galaxy as a Seyfert. Beyond a redshift of 0.1, four of the eight Seyfert 1s, 15 of the 19 Seyfert 2s, and two of the four LINERs and non-Seyferts in our sample would be misclassified as normal star-forming galaxies. Beyond a redshift of 0.3, the only galaxies recognized are the least extended or the most luminous Seyferts, with log(\Lha) greater than 41.5. This is consistent with the conclusion of \citet{peterson2006}, who made artificial Chandra observations at $z = 0.3$ and found that a sample of nearby AGN would appear optically quiescent in deep surveys. \citet{cardamone2007} suggested that a combination of observational factors, including host galaxy dilution, signal-to-noise ratio, and wavelength coverage, are responsible for hiding the nuclear emission lines of Seyfert 2s at large distances. We found that host galaxy dilution alone is enough to account for this effect. Even when a Seyfert is correctly identified as an emission-line object, dilution by H~\textsc{ii} regions may cause the galaxy to be misclassified as a normal star-forming spiral.

\section{Conclusions}

From our star formation rate comparisons, we found that the SFRs derived from the extended \Ha\ and total far-IR luminosities agreed reasonably well, although the scatter can be substantial. Comparison with the extended PAH SFR calibration showed that these relations also agree reasonably well, albeit with some scatter. The galaxies in our sample show a higher 1.4~GHz luminosity than can be attributed to star formation alone, presumably due to the presence of the Seyfert nucleus.

Comparison of nuclear \Ha\ luminosity (as best as our seeing-limited data can separate it out) with that from the host galaxy shows a correlation between the two quantities. Although selection effects might account for the absence of galaxies with high extended luminosity and low nuclear luminosity in our sample, they do not explain the absence of galaxies with high nuclear luminosity and low star formation rates. This supports the apparent correlation between AGN activity and star formation.

We determined that most of the galaxies in our sample would be identified as emission-line objects if observed at high redshift; however, whether the galaxy would be identified as a Seyfert at high redshift depends on the relative line contributions from the AGN and from H~\textsc{ii} regions. Higher-redshift Seyferts may be misclassified as normal star-forming spirals due to contamination from the host galaxy. This likely leads to incompleteness in Seyfert spectroscopic surveys beyond the local universe. We determined that beyond a redshift of 0.3, only the most luminous Seyferts in our sample would be recognized. The misclassification of distant Seyferts will be exacerbated by the fact that at higher redshifts, most host galaxies will have much higher star formation rates than they do currently, which will produce emission lines that can more easily outshine those of the active nucleus. It may be possible to overcome this problem, and identify the line emission produced by the AGN with spectroscopy at higher spatial resolution, as is possible, for example, with Hubble Space Telescope \citep[\eg][]{trump2011,atek2010}.

\acknowledgements

We gratefully acknowledge the expert assistance we have received from the members of the staff at Lick Observatory, including Elinor Gates and Paul Lyman.

This research has made use of the NASA/IPAC Extragalactic Database (NED) which is operated by the Jet Propulsion Laboratory, California Institute of Technology, under contract with the National Aeronautics and Space Administration.

Funding for SDSS-III has been provided by the Alfred P. Sloan Foundation, the Participating Institutions, the National Science Foundation, and the U.S. Department of Energy Office of Science. The SDSS-III web site is \url{http://www.sdss3.org/}.

SDSS-III is managed by the Astrophysical Research Consortium for the Participating Institutions of the SDSS-III Collaboration including the University of Arizona, the Brazilian Participation Group, Brookhaven National Laboratory, Carnegie Mellon University, University of Florida, the French Participation Group, the German Participation Group, Harvard University, the Instituto de Astrofisica de Canarias, the Michigan State/Notre Dame/JINA Participation Group, Johns Hopkins University, Lawrence Berkeley National Laboratory, Max Planck Institute for Astrophysics, Max Planck Institute for Extraterrestrial Physics, New Mexico State University, New York University, Ohio State University, Pennsylvania State University, University of Portsmouth, Princeton University, the Spanish Participation Group, University of Tokyo, University of Utah, Vanderbilt University, University of Virginia, University of Washington, and Yale University.

Research at Lick Observatory is partially supported by a generous gift from Google.

\end{document}